\documentclass[preprint]{aastex}
\usepackage{natbib,ifthen}
\usepackage{graphicx}
\usepackage{lscape}
\usepackage{amssymb,latexsym}
\usepackage{url}
\usepackage[latin1]{inputenc}

\slugcomment{}
\newcommand {\mic} {$\mu$m}
\newcommand {\ben} {\begin{eqnarray}}
\newcommand {\een} {\end{eqnarray}}
\newcommand {\msun} {M$_\odot$}
\newcommand{\msunperyr}{M$_\odot$ yr$^{-1}$}
\newcommand{\ks}{$K_{\rm s}$}
\newcommand{\jmag}{{\it J}}
\newcommand{\hmag}{{\it H}}

\begin{document}

\title{Infrared Period-Luminosity Relations of Evolved Variable Stars in the Large Magellanic Cloud}

\author{David Riebel} 
\affil{Department of Physics and Astronomy, The Johns Hopkins University, 3400 North Charles St. Baltimore, MD 21218, USA}
\email{driebel@pha.jhu.edu}
			  
\author{Margaret Meixner}
\affil{Space Telescope Science Institute, 3700 San Martin Drive, Baltimore, MD 21218, USA}
	      
\author{Oliver Fraser} 
\affil{Department of Astronomy, University of Washington, Box 351580 Seattle, WA 98195-1580 USA}

\author{Sundar Srinivasan}
\affil{Institut d'Astrophysique de Paris, 98 bis, Boulevard Arago, 75014 Paris, France}

\author {Kem Cook}
\affil{IGPP, Lawrence Livermore National Laboratory, MS L-413, P.O. Box 808, Livermore, CA 94550, USA}

\and \author{Uma Vijh}
\affil{Ritter Astrophysical Research Center, University of Toledo, Toledo, OH 43606 USA}

\shortauthors{Riebel {\it et al.}}
\shorttitle{SAGE Evolved Stars P-L Relations}


\begin{abstract}

We combine variability information from the {\it MAssive Compact Halo Objects} (MACHO) survey of the Large Magellanic Cloud (LMC) with infrared photometry from the Spitzer Space Telescope \emph{Surveying the Agents of a Galaxy's Evolution} (SAGE) survey to create a dataset of $\sim$30\,000 variable red sources.  We photometrically classify these sources as being on the first ascent of the Red Giant Branch (RGB), or as being in one of three stages along the Asymptotic Giant Branch (AGB): oxygen-rich, carbon-rich, or highly reddened with indeterminate chemistry (``extreme" AGB candidates).  We present linear period-luminosity relationships for these sources using 8 separate infrared bands (\jmag, \hmag, \ks, $3.6$, $4.5$, $5.8$, $8.0$, and $24$~\mic) as proxies for the luminosity.  We find that the wavelength dependence of the slope of the period-luminosity relationship is different for different photometrically determined classes of AGB stars.  Stars photometrically classified as O-rich show the least variation of slope with wavelength, while dust enshrouded extreme AGB stars show a pronounced trend toward steeper slopes with increasing wavelength.  We find that O-rich AGB stars pulsating in the fundamental mode obey a period-magnitude relation with a slope of $-3.41 \pm 0.04$ when magnitude is measured in the $3.6$~\mic\ band, in contrast to C-rich AGB stars, which obey a relation of slope $-3.77 \pm 0.05$.

\end{abstract}

\section{INTRODUCTION}

As intermediate mass stars ($\sim$1--8~\msun) exhaust the helium in their cores during the late stages of stellar evolution, they begin their ascent of the Asymptotic Giant Branch (AGB).  These stars consist of an inert C/O core surrounded by concentric shells of helium and hydrogen. During the early AGB (E-AGB) phase, the evolution is driven by shell hydrogen burning. Stars are typically oxygen-rich during the E-AGB phase, with a photospheric C/O ratio $<1$. Eventually, the helium shell ignites, beginning the thermally-pulsing (TP-AGB) phase. Subsequent evolution on the AGB consists of relatively long periods of hydrogen burning, punctuated at regular intervals by brief helium shell flashes (thermal pulses) that rapidly convert the helium to carbon, dramatically increasing the star's brightness for a brief period \citep{Vassiliadis1993, Habing2003}. These thermal pulses drive large-scale convective zones in the stellar interior which can ``dredge-up" nuclear-processed material, most notably newly created carbon, to the stellar surface, leading to changes in the observable surface chemistry of the star.  This is the third dredge-up for intermediate mass stars, and the result is that for stars with $\mbox{M}\lesssim 4$~\msun, the surface C/O ratio changes to become $> 1$ at which time the star is referred to as carbon-rich.  For stars with $\mbox{M}\gtrsim 4$~\msun, the temperature at the base of the convective shell is sufficiently high ($\sim 10^{8}$~K) to burn the C into N, a process known as ``hot-bottom burning (HBB)," and the star remains O-rich.

In addition, hydrodynamic instabilities drive shock waves through the stellar interior, levitating the outer layers of the star into cooler regions where dust grains can condense.  Radiation pressure then drives the dust (and the gas to which it is collisionally coupled) into the interstellar medium (ISM).  These pulsations cause AGB stas to exhibit complicated, multi-periodic variations in brightness \citep{Fraser2005}.  In addition, this process can cause an AGB star to lose mass at rates up to $\sim10^{-3}$~\msunperyr\ \citep{van Loon1999}.  This makes AGB stars one of the dominant sources of dust in the universe, and a key component in the chemical evolution of galaxies.  Toward the end of its AGB lifetime, this mass loss can completely enshroud a star with a thick circumstellar dust shell.  These ``extreme" AGB stars are nearly undetectable in the optical, and because the light from the star is dominated by thermal emission from the circumstellar shell, cannot be classified as either O-rich or C-rich based on their near-IR photometry \citep{Cioni2006}.  Most of them are believed to be C-rich, but the brightest should be so massive that HBB has left them O-rich \citep{Matsuura2009}.  Without spectroscopic confirmation, we do not propose classifications for specific sources of this class in this paper.

The {\it Surveying the Agents of a Galaxy's Evolution} (SAGE) survey \citep{Meixner2006} is an unbiased $7^{\circ} \times 7^{\circ}$  survey of the Large Magellanic Cloud (LMC) using the IRAC (3.6, 4.5, 5.8, and 8.0~\mic) and MIPS (24, 70, and 120~\mic), instruments aboard the Spitzer Space Telescope \citep{Werner2004}, intended to trace the life-cycle of the baryonic matter in the LMC.  This wavelength regime makes SAGE extremely well suited to studies of evolved stars.  The LMC makes an ideal target for studies of stellar populations, as its high galactic latitude minimizes the foreground contamination, and its distance of $\sim$50~kpc \citep{van Leeuwen2007} makes it possible to resolve individual stars while simultaneously being able to neglect its 3-dimensional structure.

Mira variables (a subset of AGB stars) have been known to follow a linear relationship in magnitude, $\log{\rm Period}$ space for nearly a century \citep[][and references therein]{Gerasimovic1928, Feast1989}.  \citet{Wood1996} extended the known dataset of AGB variables to shorter periods and established that Mira-type variables pulsate in the fundamental mode, and that other, higher order pulsational modes exist as well.  Expanding the number of data points to $\sim$1000, \citet{Wood1999} identified 4 parallel linear sequences in period-luminosity (P-L) space, and proposed mechanisms for each.  \citet{Ita2004}, using a sample of $\sim$30\,000 variable evolved stars with period information derived from the {\it Optical Gravitational Lensing Survey} \citep[OGLE;][]{Udalski1997}, further refined these sequences (see section~\ref{sec:p-l}).  Also using OGLE data, \citet{Soszynski2004} identified a new P-L sequence in the secondary, non-dominant variation in the light curves of evolved stars, and proposed this as a means of separating stars dimmer than the tip of the red giant branch (TRGB) into stars on the first ascent of the RGB and those evolving along the AGB.  With a dataset of comparable size derived from the MACHO survey \citep{Alcock1997}, \citet{Fraser2005, Fraser2008} further investigated the multi-periodic nature of stars on the sequences identified by \citet{Wood1999}.  Using early results from the SAGE survey, \citet{Glass2009} extended work on the P-L relation in AGB stars by investigating the dependence of the calculated slope of the relation on wavelength.

Our survey combines the SAGE project's photometry archive and the MACHO variability dataset to produce the largest sample of variable evolved stars, with photometry measured in multiple near-IR wavelengths, to date.  The unprecedented size of our dataset, combined with our extensive spectral coverage, reveals new patterns and details in the relationship between luminosity, variability, and spectral energy distribution amongst evolved stars.

This paper is organized as follows:  \S~\ref{sec:data} details our source selection process, the identification of the multiple sequences observed in period-luminosity space, and the division of our sample into 6 photometrically determined categories.  \S~\ref{sec:results} presents the quantitative and qualitative effects observed within our sample.  \S~\ref{sec:disc} compares our current study to previous work in the field, and \S~\ref{sec:conclusions} summarizes the conclusions of the current work.

\section{DATA} \label{sec:data}

The SAGE survey observed the LMC in two epochs, three months apart.  From each epoch, two source lists, an archive and a catalog, were generated.  The catalog stresses reliability over completeness and requires more stringent data quality standards (higher signal to noise, less crowding in the field) for inclusion of a source than the archive.  The standards for source inclusion in both archive and catalog are detailed in \citet{Meixner2006}.  

In 2009, the SAGE team released the SAGE mosaic photometry catalog and archive to the Spitzer Science Center.  By combining data from all epochs of the SAGE survey, the mosaic photometry catalog and archive are more complete than the individual epochs, and photometric errors are reduced.  The mosaic photometry data are well documented in the SAGE Data Products Description document\footnote{\url{http://data.spitzer.caltech.edu/popular/sage/20090922_enhanced/documents/SAGEDataProductsDescription_Sep09.pdf}}.  The SAGE Mosaic Photometry Catalog contains $\sim$6.5 million sources, and the archive contains $\sim$7 million.

We extract a list of 32\,744 sources from \citet{Fraser2008}'s catalog of evolved Long Period Variables (LPVs) in the LMC with a counterpart in the SAGE Mosaic Photometry Archive.  We utilize the archive in order to maximize sample size.  We use a matching radius of 2\arcsec, and keep only the closest SAGE source to a given MACHO source.  We do not include the $\sim$11\,000 sources \citet{Fraser2008} identify as the ``one-year artifact," which have an artificial period of 365 days caused by MACHO's annual observing schedule.  The database merging process can create duplicate entries for some sources, due to slight ($\lesssim0.1$~\arcsec) shifts in position between SAGE Epoch 1, Epoch 2 and the single frame mosaic photometry. By requiring that every source in our list have a unique MACHO field.tile.sequence identifier, we find $\sim$500 of these duplicate matches in our dataset and cull them from our final list.  Based on the definitions used by \citet{Fraser2008}, we assign stars to one of 6 roughly parallel sequences in period-\ks\ space (Fig.~\ref{fig:pl_seq}).  Our definitions are detailed in Appendix~\ref{app:p_l}.


We are left with a final dataset of 30\,747 evolved stars with well-defined MACHO periods and high-quality near IR photometry from 2MASS and SAGE.  Table~\ref{tab:pops} compares the number of sources in the present study to those in \citet{Fraser2008}, \citet{Srinivasan2009}, and \citet{Vijh2009}.  Our entire dataset is available online.  Table~\ref{tab:cat} presents the photometric information for a few sources from our sample as a guide to the format of the online table.

\subsection{Period-Luminosity Sequences} \label{sec:p-l}
\citet{Wood1999} identified 5 parallel period-luminosity sequences in the MACHO dataset and proposed underlying physical mechanisms for them.  In order of increasing period, the sequences of pulsating stars were named A, B, and C, with sequences D and E exhibiting variation due to an unknown mechanism.  Sequence C was identified as pulsation in the fundamental mode, sequence B as the first and second overtones, and sequence A as the third overtone.  \citet{Kiss2003} and \citet{Ita2004} split sequence B into two sequences (C$^{\prime}$ and B) representing pulsation in the first and second overtone respectively.  \citet{Fraser2005} retained the names of sequences D and E from \citet{Wood1999}, but renamed the pulsation sequences 1--4 in order of \emph{decreasing} period, in order to easily accommodate shorter period sequences, and align the naming convention with increasing pulsation overtone.  We follow the naming convention of \citet{Fraser2005, Fraser2008} because of this parallel with theory.  Numbering the sequences in order of decreasing period means that higher sequence numbers generally correspond to higher order pulsational modes, aligning the empirical naming convention and a theoretical explanation more gracefully.  \citet{Nicholls2010} has provided convincing evidence that sequence E consists of ellipsoidal binaries, systems in which the red giant member has filled its Roche lobe.  Recent work \citep[e.g][]{Nie2010,Nicholls2009} has demonstrated that the variability seen on sequence D cannot be due to binarity, but its actual cause remains unknown.    We keep the alphabetic names for these sequences to separate them from the stellar pulsation sequences.  Table~\ref{tab:seq_conv} summarizes these conventions, and Figure~\ref{fig:pl_seq} illustrates these sequences.  The precise definitions of these sequence are discussed in Appendix~\ref{app:p_l}.

\subsection{Identification and Classification of AGB Candidates} \label{sec:class}
Before being photometrically classified, all sources were de-reddened in the \jmag, \hmag\ and \ks\ bands following \citet{Glass1999}.  Specifically, we use corrections of $A_{J} = 0.112$, $A_{H} = 0.065$, and $A_{K} = 0.037$~mag.  We define AGB candidates using 2MASS and SAGE photometry \citep{Cioni2006, Blum2006, Srinivasan2009}.  We identify a source as an AGB candidate by requiring that
\begin{equation} 
K_{\rm s} > -13.333 \times (J-K_{\rm s}) + 24.666 \quad\mbox{and}\quad \mbox{\ks} < 12.05 \label{eqn:agb_def} 
\end{equation}
We classify sources as low- or moderately-obscured oxygen-rich and carbon-rich AGB candidates based on their near-IR photometry.  Specifically, a source is classified as an O-rich candidate if it lies leftward (blueward) of the line 
\begin{equation} 
K_{\rm s} = -13.333 \times (J-K_{\rm s}) +28.4 \label{eqn:o-c_split} 
\end{equation} 
in the \ks\ vs.\ $J- \mbox{\ks}$ color-magnitude diagram (CMD), and a C-rich candidate if it lies rightward (redward) of this line \citep[fig. 1]{Cioni2006}.  Extreme AGB stars are defined using their position in the [3.6] vs.\ $J-$[3.6] CMD 
\begin{equation} 
J-[3.6] > 3.1 \quad\mbox{and}\quad [3.6] < 10.5 \label{eqn:x_def} 
\end{equation}  
or the IRAC [8.0] vs.\ [8.0]$-$[3.6] CMD if a source lacks a valid \jmag\ band magnitude.  These sources are thought to be highly evolved AGB stars, enshrouded by thick circumstellar dust shells with high rates of mass loss.  

\begin{figure}
\begin{center}
\rotatebox{0}{\includegraphics*[scale=0.5]{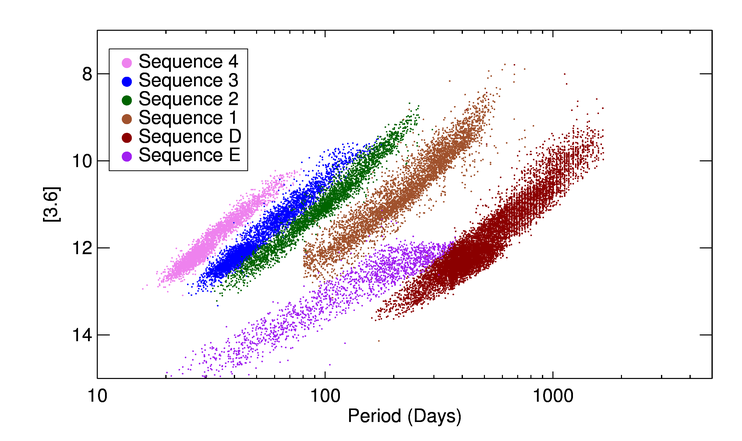}}
\caption{Period-luminosity sequences for evolved stars in the LMC.  Sequence 1 consists of stars pulsating in the fundamental mode, while sequences 2--4 are higher order pulsational modes.  Sequence E consists of ellipsoidal binary systems, and the mechanism responsible for the variation on sequence D is not known.  The naming convention follows that of \citet{Fraser2008}.  This figure is best appreciated in color, available in the online version.}
\label{fig:pl_seq}
\end{center}
\end{figure}
\begin{figure}
\begin{center}
\rotatebox{0}{\includegraphics*[scale=0.5]{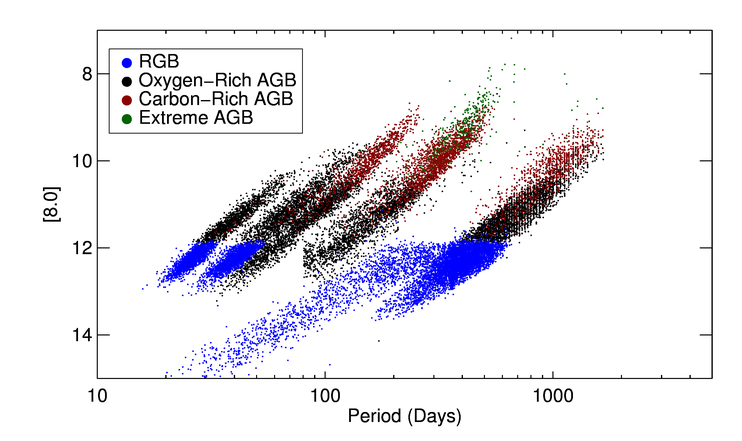}}
\caption{A reproduction of Figure~\ref{fig:pl_seq}, but color coded according to the photometrically determined classification of the source.  See \S~\ref{sec:class} for details.  This figure is best appreciated in color, available in the online version.}
\label{fig:pl_chem}
\end{center}
\end{figure}

The TRGB is located at \ks\ $= 12.0$ \citep[e.g.][]{Srinivasan2009} or {\it I} = 14.54 \citep{Cioni2006}, and is typically identified by a clear fall-off in population density \citep[e.g.][]{Nikolaev2000}.  However, while the TRGB is justified on physical grounds as a maximum luminosity for RGB stars, it does not represent a physically motivated minimum luminosity for stars on the AGB.  Examination of a plot of \ks\ magnitude vs.\ variability period \citep[][fig. 2]{Fraser2008}, shows that the sudden decrease of population density brighter than \ks\ $= 12.0$ indicative of the TRGB only appears on Period-Luminosity sequences 3 and 4.  We thus classify a star as an RGB candidate if it is fainter than the \ks\ or {\it I} band TRGB cutoff, and it lies on either sequence 3 or 4.  Stars dimmer than the TRGB on sequences 1 and 2 are classified as O-rich AGB candidates.  \citet{Soszynski2004} proposed a means of separating AGB stars dimmer than the TRGB on sequence 3 or 4 from the dominant RGB population on these sequences.  Their method relied on analysis of non-dominant periods and a distinct P-L sequence with shorter periods than sequence 4 (this sequence would be called sequence 5 in our nomenclature).  However, we do not detect this sequence in either the dominant or secondary pulsation periods of our sample, and thus we do not employ this method.  The use of periods in addition to the dominant two periods available to us grants \citet{Soszynski2004} finer resolution in separating RGB stars from AGB candidates dimmer than the TRGB, but the consistency of our results with those of other authors gives us confidence in the broad correctness of our classification.  Figure~\ref{fig:pl_chem} reproduces Figure~\ref{fig:pl_seq}, but this time color coded according to the photometrically determined chemical classification we assign to the stars in our sample.  Table~\ref{tab:pops_me} details the number of each class of star on each P-L sequence.

\begin{figure}
\begin{center}
\rotatebox{0}{\includegraphics*[scale=0.5]{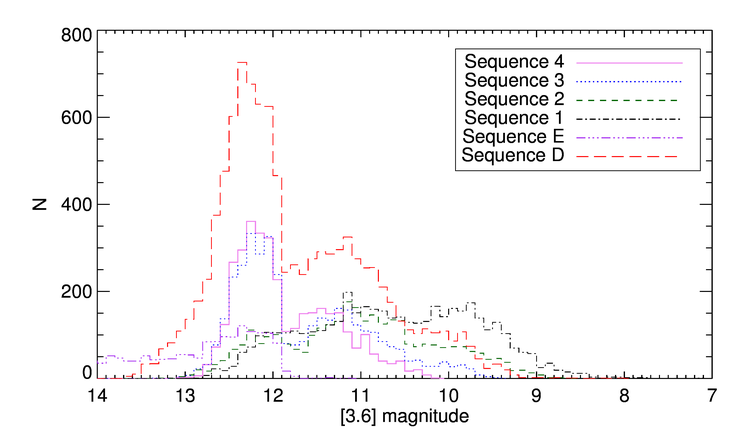}}
\caption{Histogram of [3.6] magnitude, color-coded by period-luminosity sequence.  The RGB population is clearly visible below $12^{\rm th}$ magnitude on sequences 3, 4, E and D, becoming less distinct on sequences 1 and 2.  Similarly, C-rich AGB stars, brighter than magnitude 10.5, are much more prominent on sequences 1 and 2.}
\label{fig:seq_mag_hist}
\end{center}
\end{figure}

Figure~\ref{fig:seq_mag_hist}, a histogram of [3.6] magnitude color-coded by P-L sequence, provides further justification for our classification.  Sequences 3 and 4 are dominated by a large population dimmer than $[3.6]\approx 12$.  Since there is no theoretical expectation for AGB stars to pile up at the TRGB, we interpret this as evidence for a large RGB population on Sequences 3 and 4.  Sequences 1 and 2 do show evidence of a distinct population of stars at this brightness (more pronounced in sequence 2), but neither shows the dramatic RGB population found in sequences 3 and 4.  We interpret this as a small amount of RGB contamination in sequences 1 \&\ 2.  All four sequences show a stellar population with a peak at $[3.6] \approx 11.2$, coincident with our photometrically determined O-rich AGB candidate population.  In addition, the stars photometrically classified as C-rich stars are visible in sequences 1 and 2 as a peak in the magnitude distribution brighter than $[3.6]\approx 10$.  This population is also visible in sequence 3 but absent from sequence 4.  These trends in the luminosity functions of the P-L sequences mirror and lend support to the photometrically determined chemical classifications detailed in Table~\ref{tab:pops_me}.

Highly evolved AGB stars and young stellar objects (YSOs) have similar mid-IR colors.  \citet{Whitney2008} suggest that the region of the [8.0] vs.\ [8.0]$-$[24] CMD defined by $$[8.0]-[24] > 2.2 \quad\mbox{and}\quad [8.0] > 11-1.33 \times ([8.0]-[24])$$ constitutes a region in color-magnitude space dominated by YSOs.  As all of our sources have well determined MACHO periods, none of our sample are YSOs.  Of our $\sim$30\,000 stars, we find that only 150 sources (109 RGB, 3 extreme AGB and $\sim$40 O-rich and C-rich AGB stars) fall in this region.  We note that the optical wavelengths used by the MACHO survey biases our dataset to under-sample this region of CMD space, but the extremely small fraction of our sample that falls into this region lends support to the criteria proposed by \citet{Whitney2008}.


Our sample consists of 17\,059 AGB candidate stars: 12\,172 sources photometrically classified as O-rich, 4455 as C-rich, and 432 as extreme AGB candidates.  Based on near-IR photometry and variability period, 13\,688 sources are classified as stars on the RGB.  

The \ks\ magnitude is often used as a proxy for luminosity for AGB stars \citep[e.g.][]{Wood1999,Cioni2006}. The inclusion of Spitzer IRAC and MIPS data at wavelengths out to 24~\mic\ reveals new features in the period magnitude diagram, especially among the reddest sources, the extreme AGBs, whose SEDs peak at 3~\mic\ or redward \citep{Vijh2009}.  Fig~\ref{fig:4band} compares the 2MASS \ks\ band, the IRAC $4.5$, $5.8$, and $8.0$~\mic\ bands, along with the MIPS $24$~\mic\ band as luminosity proxies, all plotted against period as in Figs.~\ref{fig:pl_seq}~\&~\ref{fig:pl_chem}.

The \ks\ band, in the upper left, is a reasonable approximation for the brightness of the low-obscuration RGB, O-rich and C-rich candidates, but comparison with the other panels in Fig.~\ref{fig:4band}, shows that this approximation fails for the extreme AGB candidates.  The RGB, O-rich and C-rich candidates appear in roughly the same positions, relative to to one another, in the 2MASS and IRAC bands.  The extreme sources, however, clearly have their luminosity underestimated by the \ks\ band.  These sources, the most luminous in our sample \citep{Srinivasan2009}, appear no more luminous than the C-rich candidates at this wavelength.  The IRAC bands, on the other hand, correctly place the extreme candidates at the brightest end of sequence 1, the fundamental mode pulsators.  

The 24~\mic\ band is sensitive to emission from cool dusty envelopes. The observed flux from the sources surrounded by the most optically thick dust shells is dominated by the emission from their outermost, coolest layers. For this reason, they are able to produce significant 24~\mic\ fluxes \citep[see Fig. 10,][]{Srinivasan2009}. The RGB/AGB stars with optically thin dust shells are either faint or undetected at 24~\mic\ (for example, $\lesssim 3$\% of the RGB stars in our sample have valid measured 24~\mic\ fluxes) because their SEDs peak at much shorter wavelengths. The 24~\mic\ flux is therefore not representative of the intrinsic brightness of these sources. This poor estimate for intrinsic brightness causes the sequence structure to break down entirely.

\section{RESULTS}\label{sec:results}

\subsection{Period--Magnitude Relations}\label{sec:pl_rel}
Using the method of weighted least-squares, linear models of the form \begin{equation} m = A \times \log{P} + B \end{equation}\label{eq:pl} were fit to each of the sequences as a whole, and to each subgroup (O-rich, etc.) within each sequence.  Periods and amplitudes are from the dominant mode of variability derived by \citet{Fraser2005,Fraser2008} from MACHO blue-band lightcurves.  Periods were considered free of errors, and because the SAGE observations are not phase corrected, magnitudes were weighted by the inverse of the quadrature sum of the 1-$\sigma$ photometric error and one-half the MACHO peak-to-peak variability amplitude.  Table~\ref{tab:fit} lists the linear fit parameters thus determined for each of the 6 sequences, using the IRAC [3.6] magnitude as the luminosity proxy.

Previous work has indicated that Mira variables exhibit much lower amplitude variation in the IR than in the optical \citep[Lattanzio \& Wood (2003), in:][]{Habing2003}, and thus using one-half of the MACHO amplitude in the error term might overestimate the true scatter in the data due to intrinsic variation.  To investigate this, we extracted a sample of $\sim$100 sources from the catalog of \citet{Vijh2009} with well-determined MACHO periods between 150 and 200 days.  At these periods, the three month cadence of the SAGE observations could sample these stars anywhere between maximum and minimum brightness.  Thus the range of variation in brightness between Epoch 1 and 2 observations could represent the full mid-IR peak to peak amplitude of variation.  The MACHO amplitudes for the O-rich and C-rich AGB candidates are statistically very similar to the variation seen in this smaller sample.  The small sample from \citet{Vijh2009} has a median mid-IR variation of $\sim$0.2~mag between SAGE Epoch 1 and Epoch 2.  The median optical amplitudes for all O-rich and C-rich sources together are 0.2 and 0.3 magnitudes respectively.  In addition, we compared the distribution of the MACHO variability amplitudes to the distribution of the residual scatter (Table~\ref{tab:fit}) about the P-L fits.  The residual scatter of $\sim$0.5 magnitude about the best fit line (see Table~\ref{tab:fit}) is due to the intrinsic variability of the sources.  The SAGE mosaic photometry is constructed from the two epochs of the SAGE survey, but lacks the time-stamp information necessary for the phase-correction necessary to compute a true mean magnitude for these variable sources.  Some of our sources have been observed near maximum brightness, while other were near their minimum brightness.  The residual scatter about the P-L relations gives us an indication of the scale of the IR variability of our sample.  Both quantities followed similar distributions, and were of the same scale.  This justifies our use of the MACHO optical variability amplitude as a measure of the IR variability of our sources for the estimation of the uncertainty in the IR mean magnitudes in the P-L relations.

Figure~\ref{fig:seq1} shows the linear fits to the stars in sequence 1 in the [3.6] vs.\ period diagram.  Sequence 1 is consistent with stellar pulsation in the fundamental mode \citep{Fraser2005}, and was chosen because it is the only sequence which contains a significant number of detected extreme AGB candidates.  The $3.6$~\mic\ IRAC band was chosen because it is the most sensitive of the SAGE bands \citep{Meixner2006}.  The different trends followed by the different types of evolved star candidates is apparent.  The plotted linear fits are:

\begin{tabular}{lc}
	O-rich & $[3.6] = (-3.41 \pm 0.04) \times \log{P} + (18.88 \pm 0.09)$ \\
	C-rich & $[3.6] = (-3.76 \pm 0.05) \times \log{P} + (19.35 \pm 0.12)$ \\
	Extreme AGB & $[3.6] = (-4.27 \pm 0.19) \times \log{P} + (20.4 \pm 0.49)$
\end{tabular} 

The increasing steepness of the linear relations followed by the different categories of AGB star would be consistent with a non-linear P-L relation if the evolved variable stars were not divided into sub-categories.  We experimented with fitting quadratic functions to all the stars in each sequence, regardless of chemical classification.  While we found such fits to be consistent with our data, we did not find them to be statistically superior fits to the linear models presented here.

Table~\ref{tab:fit} lists the parameters of the linear fits to each of the 6 sequences in the $3.6$~\mic\ band.  Identical tables for all of the 2MASS, IRAC and MIPS bands used in this work are included as Appendix~\ref{app:fits}, in the online edition.

Figure~\ref{fig:1_comp} plots the slope of the linear fit to the sources of each type of evolved star, in sequences 1 and 2, in all the IR bands.  Because we only use the period measured from the light curve in the MACHO blue band, this is a fixed value for all of our sources.  The slope of the $\log{P}$-magnitude relation is therefore a measure of how different the brightness in a particular band is between the longest period stars of a particular type and the shorter period stars of the same type on the same sequence. Figure~\ref{fig:1_comp} shows that O-rich AGB candidates follow roughly the same linear relationship in all IR bands, showing that there is little difference between the SEDs of the O-rich AGB candidates with the longest periods, and those with the shortest periods.  The extreme AGB candidates, on the other hand, show a very strong trend toward steeper slopes as one moves towards longer wavelengths.  The slope of the P-L relation is a measure of the dependence of brightness in a certain band to period.  Smaller values for slope indicate smaller differences in brightness between short and long period stars within one class on one sequence (e.g. RGB stars on sequence 3).  Steep slopes indicate large differences between short and long period stars within one class on one sequence (e.g. extreme AGB stars on sequence 1).  All extreme AGB candidates become brighter as one looks at redder bands because they possess dusty circumstellar envelopes which reprocess the near-IR stellar photospheric emission to longer IR radiation.  However, this mid-IR brightening is more dramatic for the longest period sources on a given P-L sequence, which causes the slope to become steeper and indicates that the longest period sources have the thickest circumstellar dust shells.  Thus, the steep slopes of the P-L relations followed by extreme AGB candidates at longer wavelengths indicate not just the presence of circumstellar matter, but show that period is more closely tied to the presence of circumstellar matter in extreme AGB stars than in, e.g. O-rich AGB stars.

The slope of the $\log{\rm P}$-magnitude relation for C-rich AGB stars shows a more complicated variation with wavelength.  In the IRAC [5.8] band (and to a lesser extent in the $4.5$~\mic\ band), the slope becomes noticeably less steep, indicating that there is less brightness contrast between the C-rich stars with longest periods and those with shorter periods on the same sequence in this band.  The spectra of C-rich stars have a strong CO absorption feature at $\sim$4--5~\mic\ \citep[e.g.][fig.~4]{Aringer2009}.  The observed decrease in the slope of the $\log{\rm P}$-magnitude relation in the 5.8~\mic\ band compared to neighboring bands indicates a possible relationship between the strength of this feature and the pulsation period of carbon-rich stars.  Specifically, that longer-period C-rich AGB stars may have stronger CO absorption features than those stars with shorter periods.  While many of them are thought to be carbon-rich, extreme AGB candidates do not show this effect because their IR emission is dominated by their circumstellar dust shells.  The absence of this effect in O-rich AGB candidates, which also feature CO in their photospheres, argues against our hypothesis.  Further IR spectroscopic follow-up, such as the SAGE-SPEC survey \citep{Kemper2010} will shed further light on this.

Figure~\ref{fig:1_comp} also has implications for future observational strategies.  AGB stars are a dominant source of IR light in galaxies with intermediate- and old-age stellar populations.  With major IR observatories such as the \emph{James Webb Space Telescope} coming online in the next few years, the AGB period-luminosity relationship constitutes a possible distance indicator.  Figure~\ref{fig:1_comp} shows that the observed scatter in the P-L relation could be reduced, while maximizing sample size, by observing in the near-IR, specifically the \ks\ band ($\sim$2.2~\mic).  In this band, all AGB candidates brighter than the TRGB obey very nearly the same P-L relation.  Furthermore, the slope of the relation measured for AGB candidates in the \ks\ band will be the slope of the relation followed by O-rich AGB stars (which can be photometrically classified) at all IR wavelengths out to 8~\mic\ (Also see Table~\ref{tab:kfit}).  Figure~\ref{fig:seds} plots the median SED for each classification of star used in this paper.  Error bars represent the spread of each flux within the population, not measurement uncertainty.  The \ks\ band is the location of the peak of the SED for the majority of C-rich AGB stars \citep[also see][]{Srinivasan2009}, and is very near the peak for both the O-rich AGB stars and those on the first ascent of the RGB.  Furthermore, 97\%\ of our sample (including 82\%\ of the extreme AGB stars) have a valid flux measured in the \ks\ band, implying that this band will provide an adequate sample size.  The intrinsic brightness of the reddest, most evolved AGB stars will be severely underestimated (see \S~\ref{sec:class}), but these stars can be identified by near-IR color and excluded from the sample to reduce scatter in the derived relationship.  Table~\ref{tab:kfit} presents all the derived \ks\ vs. $\log{\rm P}$ relations for our sample.

\subsection{Long Secondary Periods} \label{sec:lsp}

Approximately 30\%\ of variable AGB stars exhibit long secondary periods (LSPs), variations which occur on time scales nearly an order of magnitude longer than pulsation in the fundamental mode (Sequence 1).  The LSP phenomenon is represented in our data set by sequence D, and the mechanism behind it is still unknown \citep{Nicholls2009,Nie2010}.  \citet{Fraser2008} noted that the variability properties of sequence D are consistent with a population drawn from all of the other P-L sequences.  The LSP is known to be related to mass-loss \citep{Wood2009}, but in an as yet unknown manner.

The lower left panel of figure~\ref{fig:allseq_comp} shows the variation of the LSP with wavelength.  The slopes of the P-L relations followed by the AGB candidate stars on sequence D show the same dependence on wavelength as the stars on the other sequences.  O-rich AGB stars show little variation in P-L relation slope with wavelength, and C-rich AGB stars show identical behavior in Sequences 1, 2, 3 and D.  We do not include the Extreme AGB candidates from sequence D in Figure~\ref{fig:allseq_comp} because the fits, based on only 18 stars, are not statistically significant (Table~\ref{tab:kfit}).  Whatever mechanism lies behind the LSP, it connects the period of variation of the star with its SED in the same manner as fundamental and overtone stellar pulsation.





\subsection{RGB P-L relation}\label{sec:rgb_pl}

Figure~\ref{fig:allseq_comp} shows how the slopes of the RGB P-L relations on sequences 3, 4, E and D relate to those determined for the various classes of AGB candidates.  The slope of the RGB-star P-L relation is relatively wavelength-independent (qualitatively similar to the behavior seen in O-rich candidates) and is consistently less steep than the AGB-star slope at all wavelengths.  This effect has been noted by other authors \citep{Ita2004,Glass2009}, and the persistence of this trend at longer wavelengths lends support to the conclusion that the populations above and below the TRGB on these sequences are indeed dominated by different classes of star.

Stellar evolutionary models \citep[e.g.][]{Castellani2003} predict (for a limited mass range) an offset in variability period at a given luminosity between AGB stars and those on the first ascent of the RGB of $\delta \log{P} \approx 0.03$, with RGB stars having the longer period.  \citet{Kiss2003} observed this effect in their OGLE based survey, and we confirm it using MACHO.  This effect is not immediately visually apparent in our sample due to crowding (Figure~\ref{fig:pl_chem}).  However, by solving our liner P-L relationships for $\log{P}$ at a fixed [3.6] = 12.0, we find $\delta \log{P} = 0.05$ between the stars we define as RGB candidates and those classified as AGBs on sequences 3, 4, and D.

\citet{Nicholls2010} examined the orbital parameters of a sample of sequence E binary stars, and demonstrated that their sample did indeed  consist of ellipsoidal binary systems, in contrast to a sample of field red giants and a sample of stars from sequence D \citep{Nicholls2009}.  In addition, they looked for, but did not find, evidence of an IR excess in the colors of sequence E stars, an indication of mass loss.  Figure~\ref{fig:allseq_comp} shows that the P-L relations of the RGB stars on Sequence E and D behave, as a function of wavelength, exactly like the the RGB stars on sequences 3 and 4.  The lack of change in the slope of the P-L relation indicates that period is not coupled to mass loss in these stars.

The P-L relations of RGB stars on sequences E \&\ D are systematically less steep than those of the RGB stars on the pulsational sequences 3 and 4. The slope may be less steep because of the photometric definition of an RGB star used in this study, see Appendix~\ref{app:p_l}.  Nevertheless, we confirm the null result of \citet{Nicholls2010} in that we see no evidence of any difference in the RGB P-L relations as a function of wavelength for sequence E \&\ D compared to those of sequence 3 and 4.

\section{DISCUSSION} \label{sec:disc}
\subsection{Comparison to Previous Work}
The period-luminosity relationship for AGB stars has been examined by many authors over the past several decades \citep[e.g.][]{Feast1989, Hughes1990, Wood1999, Glass2009}.  Previous studies have often relied on a few tens to a few hundreds of evolved LPVs with brightnesses often measured in the \ks\ band.  The present work is notable for having the largest sample of LPVs with photometric data measured in the most IR bands to date.  This has allowed us to reduce the uncertainties in these relations by a factor of $\sim$5.

\subsubsection{Comparison to \citet{Feast1989}} \label{sec:feast}
\citet{Feast1989} observed 49 evolved LPVs in the LMC and constructed P-L relations in the \jmag, \hmag\ and \ks\ bands, as well as bolometric magnitudes, which were used in several subsequent studies \citep[e.g.][]{Vassiliadis1993, Whitelock2003}.  In the \ks\ band, they find a slope of $-3.47 \pm 0.19$ fits the O-rich stars in their sample, consistent with our slope of $-3.31 \pm 0.04$.   Our fit to the C-rich AGB candidates in the \ks\ band is also consistent with their derived relation.  We find a slope of $-3.16 \pm 0.04$ compared to their value of $-3.30 \pm 0.40$.  They do not consider their fits to C-rich Miras in the \jmag\ and \hmag\ bands to be useful.


\subsubsection{Comparison to \citet{Ita2004}}\label{sec:ita}
\citet{Ita2004} used variability information from the OGLE survey of the LMC \citep{Udalski1997} to assemble a dataset of 35\,000 variable stars in the LMC.  This sample size allowed them to confirm the existence of a distinct population of variable stars below the TRGB, concentrated on sequences 3 and 4 (their sequences A and B) first noted by \citet{Kiss2003}.  They reserve judgment on the interpretation, put forth by \citet{Kiss2003} and used in this paper, that this population consists of a large number of first-ascent RGB stars with some AGB contamination.  

Most of the \ks\ P-L relations fits derived in our study (Table~\ref{tab:kfit}) are consistent with those in \citet{Ita2004}.  They obtain a slope of of $-3.369 \pm 0.099$ for the C-rich stars on sequence 1, consistent with our slope (\S~\ref{sec:feast}).  Our most discrepant fits are those to stars brighter than the TRGB in sequences 3 and 4.  We find a slope of $-3.73 \pm 0.04$ for the 1434 O-rich AGB stars we identify on sequence 4.  \citet{Ita2004} do not differentiate between O-rich and C-rich AGB stars here, but they find a slope of $-3.289 \pm 0.047$ to all 510 stars brighter than the TRGB on this sequence.  Similarly, on sequence 3 our O-rich AGB stars are best fit with a slope of $-3.85 \pm 0.04$ compared to $-3.356 \pm 0.052$ in \citet{Ita2004}.  This is most likely due to slight differences in sequence definition.  Sequence 3 is almost blended with sequence 2, and precisely where the dividing line between these two sequences is drawn can greatly impact the derived fit.

Our study provides a valuable complement to that of \citet{Ita2004}, due to the fact that our sample is of comparable size but is derived from completely independent observations.

\subsubsection{Comparison to \citet{Glass2009}} \label{sec:glass}
\citet{Glass2009} did their own reduction of a subset of the SAGE and MACHO datasets to compute P-L relationships for AGB stars in multiple bands in the IR.  Our current work differs most noticeably in sample size, their sample consisting of $\sim$1800 stars and ours $\sim$30\,000.  Precise comparison of our derived P-L relations is complicated, however, by several differences in how we categorize our datasets.  The sample of \citet{Glass2009} is not large enough to distinguish what we identify as sequences 2 \&\ 3, combining them into a single sequence B.  See Table~\ref{tab:seq_conv} for the different naming conventions used.  Finally, \citet{Glass2009} only fit P-L relations to stars they classify as O-rich, defined as a star with $(J- \mbox{\ks}) < 1.6$, a redder definition than our own (detailed in section~\ref{sec:class}).  The reddest star that we classify as O-rich has $(J-\mbox{\ks} ) = 1.4$.  

With these differences in mind, our results show reasonable agreement with those of \citet{Glass2009}.  On sequence 1 for example, our fits for the AGB stars are consistent within the errors in nearly every band from \ks\ to 8~\mic.  The notable exception to this broad consistency is in the IRAC [5.8] band on sequence 1, which we discuss above (\S~\ref{sec:pl_rel}) as having an anomalously less steep P-L relation.  This effect is not observed in \citet{Glass2009}, a result we attribute to sample size.  Their fit is based on 41 AGB stars on sequence 1, whereas ours is based on 1812.

\subsubsection{Comparison to \citet{Vijh2009}} \label{sec:uma}
By comparing epoch 1 and epoch 2 fluxes from the SAGE point source catalog, \citet{Vijh2009} identified 1967 variable sources in the LMC.  Having only two data points, they were not able to determine specific variability parameters such as period or amplitude for these sources.  We identify 731 of these sources in the SAGE single frame mosaic photometry archive with MACHO detections.  Of the remaining sources, with no MACHO counterpart, \citet{Vijh2009} identified 41 as O-rich, 60 as C-rich, and 540 as extreme AGB stars.  Only 1.5\% of our sample is classified as extreme.  On the other hand, 84\% of the AGB variables identified by \citet{Vijh2009} without a MACHO counterpart are classified as extreme AGB stars.  By requiring our sources to have a MACHO detection, this effect is expected.  The MACHO survey focused on the bar of the LMC, and extreme AGB candidates are not as concentrated in the bar as are O- and C-rich AGB stars \citep[][fig. 1]{Blum2006}.  Furthermore, because the MACHO survey used two optical bandpasses, we expect our sample to miss the reddest, most enshrouded evolved stars in the SAGE catalog, which are not visible in the optical.  The variables identified by \citet{Vijh2009} are indeed systematically redder than the sample identified here.  Figure~\ref{fig:u_cmd} presents a CMD of all the AGB candidate sources in this paper, using the SAGE [3.6] and [8.0] bands.  The variables detected by \citet{Vijh2009} are overlaid as large diamonds.  Figure~\ref{fig:u_hist} presents a histogram comparing the [3.6]--[8.0] colors of the extreme sources in our study with those of \citet{Vijh2009}.

\subsubsection{Summary of Comparison to Other Surveys}\label{sec:comparison_summary}
Overall, our quantitative results are consistent with previous studies of the P-L relation in AGB stars.  Differences that exist are small, and can be attributed to differences in sample size.  Our current work exhibits two primary strengths, sample size and wavelength coverage.  Consisting of a sample of 30\,000 evolved stars with magnitudes measured in 8 different near-IR bands, our sample is far larger than most previous studies.  \citet{Ita2004} use a similar sized dataset, but lack the wavelength coverage the SAGE survey gives this work.  Similarly, \citet{Glass2009} also use the SAGE catalog to investigate the dependence of the P-L relation on wavelength, but use only a small fraction of the full SAGE source list.  We combine the individual strengths of these two works to produce a comprehensive view of evolved star variability in the IR.

\citet{Vijh2009} also utilized the entire SAGE catalog to probe variability.  We expand on this work with the inclusion of the MACHO catalog, which allows us to precisely determine the period and amplitude of the variation of these stars.  However, this additional information introduces two selection biases to our sample that do not affect the Vijh sample.  The MACHO survey was conducted in two non-standard optical filters.  By requiring that all of our sources have valid MACHO detections, we miss the reddest of the evolved stars.  In addition, the MACHO survey focused on the bar of the LMC, and this spatial bias also affects our sample.

\subsection{Comparison to Models of AGB Evolution}\label{sec:evolution}
Figure~\ref{fig:pl_chem} reproduces the [3.6] vs.\ period plot of Figure~\ref{fig:pl_seq}, but this time color coded according to RGB, O-rich, C-rich, or extreme AGB classification.  By grouping the sources according to this classification, we see trends broadly consistent with synthetic AGB evolution codes \citep[e.g.][]{Vassiliadis1993}, which predict that the structural and chemical changes during the star's evolution along the AGB should be accompanied by a general trend toward slightly higher luminosities and longer pulsational periods.

Variable stars on their first ascent of the RGB are dimmer than the AGB stars at later stages of evolution, and highly concentrated on sequences 3 and 4, the shortest period sequences.  Oxygen-rich AGB candidates are distributed across all 4 pulsational sequences.  AGB candidates photometrically classified as carbon-rich are also found on all sequences, but are much more concentrated on the long period sequences (only 29 are on sequence 4, compared to $\sim$2000 on sequence 1).  Extreme AGB stars, which are heavily enshrouded by circumstellar dust, are found in significant numbers only on sequence 1, the longest period sequence.  Table~\ref{tab:pops_me} details the number of each sub-class of star on each sequence.  We see clear trends toward brighter magnitudes and longer periods as one moves from RGB candidates to O-rich, C-rich, and finally extreme AGB candidates.  Figures~\ref{fig:mag_hist} \& \ref{fig:per_hist} illustrate the distributions of our sample in period and [3.6] magnitude.  Previous studies \citep[e.g.][]{Feast1989, Whitelock2003}, have noticed a dramatic steepening of the AGB P-L relation at $\sim$420~days.  Figure~\ref{fig:per_hist} shows that this is precisely the period at which extreme AGB candidates come to dominate over all other classes of stars.  Note that in figure~\ref{fig:per_hist}, the extreme AGB candidate graph has been magnified to be visible compared to the much larger O-rich and C-rich AGB populations.  Extreme AGB candidates do dominate at periods greater than 420 days, but the effect is smaller than figure~\ref{fig:per_hist} shows.  As we discuss in \S~\ref{sec:pl_rel}, we find that this class of star does obey a steeper P-L relation.  


The slope of the P-L relation is a measure of the brightness contrast between stars with long periods and those of the same type with shorter periods.  If the P-L relation becomes increasingly steeper with longer wavelength, it is an indication that longer-period stars are emitting more of their energy at those wavelengths, an indication of a greater amount of circumstellar matter re-processing the star's light to the red.  More circumstellar matter implies a higher rate of mass loss \citep[][and references therein]{Srinivasan2009}.  Thus, we interpret the lack of a relationship between P-L slope and wavelength amongst RGB stars and O-rich AGB stars to indicate that pulsation period does not strongly impact mass-loss rates at these stages of stellar evolution.  However, amongst heavily enshrouded extreme AGB stars, there is a stronger coupling between the two.

Other authors \citep[e.g.][]{Schultheis2004, Glass2009} have examined plots of color vs.\ period in order to ascertain a period at which significant reddening indicates the onset of mass loss.  Our determination of the slope of the P-L relation at multiple wavelengths contains and quantifies these other studies.  Quantitative period-color relations can be easily derived from the fits given in this paper.  Given P-L fits in two bands, $m_{1}$ and $m_{2}$, we obtain the color-period relation $(m_{1} - m_{2}) = (A_{1} - A{2})*\log{\rm P} +(B_{1} - B_{2})$ with A and B defined as in Eqn.~\ref{eq:pl}.  \citet{Groenewegen2006} suggests the [3.6] $-$ [4.5] color as a good indicator of the mass-loss rate from AGB stars.  Through the period-color relations that can be derived from this work, such proxies can be readily connected to variability parameters, and the results compared to the predictions of theoretical models of AGB evolution.  We intend to investigate the connections between variability and mass-loss rate using the quantitative predictions of such models in future papers.

\section{Conclusions} \label{sec:conclusions}
We present the largest multi-wavelength IR investigation of the period-luminosity relationship of evolved stars in the LMC to date.  We find that the slope of the P-L relation followed by O-rich AGB stars and stars on the first ascent of the RGB is relatively independent of wavelength.  In contrast, we find a strong dependence of slope of the P-L relation with wavelength amongst the most evolved, most enshrouded, extreme AGB stars.  C-rich AGB stars, which represent an intermediate stage of evolution along the AGB, show a slight dependence of slope on wavelength, with a possible correlation between variability period and the strength of the 5~\mic\ CO absorption feature.  

We find that for accurate characterization of the IR P-L relationship for AGB stars, the \ks\ band is the best choice.  It offers a superior combination of sample size; low scatter due to the fact the both O-rich and C-rich AGB stars obey identical P-L relations when measured in this band; and further applicability, in that once the P-L relation is determined in the \ks\ band, the slope of the relation for O-rich AGB stars and RGB stars has been determined at all near-IR wavelengths.  Slopes of P-L relations for evolved stars in the \ks\ band can be found in Table~\ref{tab:kfit}.

There is a systematic trend towards steeper slopes in the magnitude vs. $\log {\rm P}$ relation as one looks at categories of stars at later and later stages of stellar evolution.  This trend would be consistent with a non-linear P-L relation if all variable evolved stars were examined together, without photometrically determined classifications.

\acknowledgments
The authors would like to thank Bernie Shiao for his invaluable assistance with the SAGE database.  In addition, the courteous and helpful comments of the anonymous referee were invaluable in the completion of this document.

The SAGE Project is supported by NASA/Spitzer grant 1275598 and NASA NAG5-12595.

This paper utilizes public domain data obtained by the MACHO Project, jointly funded by the U.S. Department of Energy through the University of California, Lawrence Livermore National Laboratory under Contract no. W-7405-Eng-48, by the National Science Foundation through the Center for Particle Astrophysics of the University of California under cooperative agreement AST-8809616, and by the Mount Stromlo and Siding Spring Observatory, part of the Australian National University.

This publication makes use of data products from the Two Micron All Sky Survey, which is a joint project of the University of Massachusetts and the Infrared Processing and Analysis Center/California Institute of Technology, funded by the National Aeronautics and Space Administration and the National Science Foundation.

\begin{deluxetable}{lcccccc}
\tabletypesize{\scriptsize}
\tablecaption{Dataset Populations}
\tablewidth{0pt}
\tablehead{\colhead{Source} & \colhead{O-rich} & \colhead{C-rich} & \colhead{Extreme AGB stars} &  \colhead{Total AGB stars} & \colhead{RGB stars} & \colhead{Total Sources} }
\startdata
\citet{Fraser2008}\tablenotemark{a}      &  N/A      & N/A   & N/A   & N/A      &  N/A     & 56\,453   \\
\citet{Srinivasan2009}\tablenotemark{b}  &  17\,958  & 5179  & 1428  & 24\,565  &  N/A     & 24\,565   \\
\citet{Vijh2009}\tablenotemark{b}        &	353      & 426   & 820   & 1599     &  N/A     & 1967      \\
This work                                &  12\,172  & 4455  & 432   & 17\,059  &  13\,688 & 30\,747   \\
\enddata
\tablecomments{Number of sources used in recent studies of AGBs and variable sources in the LMC.  Classification of sources as RGB, O-rich, etc.\ is done photometrically, using eqns. \ref{eqn:agb_def} -- \ref{eqn:x_def}}
\tablenotetext{a}{\citet{Fraser2008} did not classify their AGB candidates as O-rich, C-rich, etc.}
\tablenotetext{b}{Did not use RGB classification}
\label{tab:pops}
\end{deluxetable}

\begin{landscape}
\begin{deluxetable}{lcccccccccc}
\setlength{\tabcolsep}{0.05in}
\tablecaption{Source List of AGB Candidates}
\tabletypesize{\scriptsize}
\tablewidth{0pt}
\tablehead{
\colhead{SAGE ID\tablenotemark{a}} & \colhead{MACHO FTS\tablenotemark{b}} & \colhead{2MASS ID} & \colhead{Classification\tablenotemark{c}} & \colhead{Amplitude\tablenotemark{d}} & \colhead{Period\tablenotemark{d}} & \colhead{J mag} & \colhead{$\sigma_{\rm J}$} & \colhead{H mag} & \colhead{$\sigma_{\rm H}$} & \colhead{\ldots \tablenotemark{e}} }
\startdata
  SSTISAGEMA J050220.31-691523.5 &       1.3564.19 &    1204226 &   O &    0.121 &   512 &  12.93 &   0.02 &  12.01 &   0.03 & \ldots \\ 
  SSTISAGEMA J050324.49-684624.1 &       1.3692.24 &    1204566 &   C &    0.250 &  1350 &  12.33 &   0.05 &  11.32 &   0.04 & \ldots \\ 
  SSTISAGEMA J054643.17-703315.3 &    12.10805.486 &   12528609 &   X &    1.485 &   512 &  15.47 &   0.07 &  13.29 &   0.03 & \ldots \\ 
  SSTISAGEMA J050214.21-692048.9 &       1.3563.74 &   12528609 &   R &    0.092 &    72 &  14.58 &   0.03 &  13.85 &   0.03 & \ldots \\ 
\enddata
\tablenotetext{a}{The SAGE single frame mosaic photometry identifier, containing the coordinates of the IRAC source}
\tablenotetext{b}{The MACHO Field Tile Sequence identifier of the corresponding MACHO source}
\tablenotetext{c}{Photometric classification of source as RGB, oxygen-rich, carbon-rich, or extreme AGB candidate.  See text for details}
\tablenotetext{d}{Variability parameters from \citet{Fraser2008}}
\tablenotetext{e}{2MASS and SAGE Mosaic Archive photometry and 1-$\sigma$ errors.  The online version of this table contains all magnitudes and errors in the \jmag, \hmag, \ks, [3.6], [4.5], [5.8], [8.0], and [24] bands, for all 30\,747 sources in our sample.  A portion is shown here for guidance regarding its form and content.}
\label{tab:cat}
\end{deluxetable}
\end{landscape}

\clearpage

\begin{deluxetable}{ccccl}
\tabletypesize{\scriptsize}
\tablecaption{Period-Luminosity Sequence Conventions}
\tablewidth{0pt}
\tablehead{ \colhead{ \citet{Fraser2005} }& \colhead{ \citet{Wood1999} } & \colhead{ \citet{Ita2004}} & \colhead{ \citet{Glass2009}} & \colhead{Theoretical Explanation} }
\startdata
Sequence 1 & Sequence C  & Sequence C  & Sequence C & Pulsation in Fundamental Mode  \\
Sequence 2 & Sequence B  & Sequence C$^{\prime}$    & Sequence B & First/Second Overtone Pulsation \\
Sequence 3 & Sequence B  & Sequence B  & Sequence B & First/Second Overtone Pulsation \\
Sequence 4 & Sequence A  & Sequence A  & Sequence A & Third Overtone Pulsation\\
Sequence D & Sequence D  & Sequence D  & Sequence D & Undetermined \\
Sequence E & Sequence E  & Sequence E  & N/A & Ellipsoidal binaries \\
\enddata
\tablecomments{Naming conventions used to refer to the various period-luminosity sequences first identified by \citet{Wood1999}.  This paper follows the convention of \citet{Fraser2005}, because the numeric naming scheme more closely aligns with the theoretical explanation.}
\label{tab:seq_conv}
\end{deluxetable}

\begin{deluxetable}{lcccc}
\tabletypesize{\scriptsize}
\tablecaption{Sequence Populations}
\tablewidth{0pt}
\tablehead{Sequence & \colhead{O-rich} & \colhead{C-rich} & \colhead{Extreme} &  \colhead{RGB} }
\startdata
     Sequence 1 &   2229 &   1817 &    371 &      0 \\
     Sequence 2 &   2292 &    891 &      1 &      0 \\
     Sequence 3 &   1707 &     56 &      1 &   1992 \\
     Sequence 4 &   1434 &     29 &      0 &   2068 \\
     Sequence D &   3364 &    971 &     18 &   5611 \\
     Sequence E &      0 &      0 &      0 &   1727 \\
\enddata
\tablecomments{Populations of sources of various types on each sequence.  Amongst the stellar pulsational sequences, RGB stars are concentrated on the short period sequences (1--4), O-rich AGB candidates are equally distributed across all sequences, and C-rich and extreme AGB candidates are concentrated on the long period sequences.}
\label{tab:pops_me}
\end{deluxetable}


\begin{deluxetable}{lccccc}
\tabletypesize{\scriptsize}
\tablecaption{[3.6] vs. $\log{\rm P}$ linear fit parameters }
\tablewidth{0pt}
\tablehead{
\colhead{Classification} & \colhead{Slope} & \colhead{Intercept} & \colhead{Residual} & \colhead{r$^{2}$} & \colhead{Number} }
\startdata
&\multicolumn{4}{c}{ {\bf Sequence 4} }   & \\
RGB &  -3.67 $\pm$   0.05 &  17.45 $\pm$   0.08 & 0.127 & 0.687 &   2064 \\
O-rich &  -3.90 $\pm$   0.04 &  17.58 $\pm$   0.06 & 0.121 & 0.886 &   1432 \\
C-rich &  -4.14 $\pm$   0.36 &  18.04 $\pm$   0.58 & 0.144 & 0.829 &     29 \\
All Stars &  -4.56 $\pm$   0.02 &  18.68 $\pm$   0.03 & 0.138 & 0.935 &   3525 \\
 &\multicolumn{4}{c}{ {\bf Sequence 3}}   & \\
RGB &  -3.32 $\pm$   0.05 &  17.57 $\pm$   0.08 & 0.122 & 0.704 &   1983 \\
O-rich &  -4.09 $\pm$   0.04 &  18.62 $\pm$   0.07 & 0.182 & 0.867 &   1704 \\
C-rich &  -3.81 $\pm$   0.27 &  18.04 $\pm$   0.51 & 0.248 & 0.788 &     56 \\
All Stars &  -4.51 $\pm$   0.02 &  19.45 $\pm$   0.03 & 0.167 & 0.939 &   3744 \\
& \multicolumn{4}{c}{ {\bf Sequence 2} }  & \\
O-rich &  -4.24 $\pm$   0.02 &  19.51 $\pm$   0.04 & 0.153 & 0.942 &   2289 \\
C-rich &  -4.25 $\pm$   0.05 &  19.32 $\pm$   0.11 & 0.189 & 0.890 &    887 \\
All Stars &  -4.54 $\pm$   0.02 &  20.04 $\pm$   0.03 & 0.177 & 0.960 &   3176 \\
& \multicolumn{4}{c}{ {\bf Sequence 1} }  & \\
O-rich &  -3.41 $\pm$   0.04 &  18.89 $\pm$   0.10 & 0.271 & 0.729 &   2221 \\
C-rich &  -3.77 $\pm$   0.05 &  19.35 $\pm$   0.12 & 0.251 & 0.777 &   1813 \\
Extreme AGB &  -4.27 $\pm$   0.19 &  20.37 $\pm$   0.49 & 0.336 & 0.586 &    371 \\
All Stars &  -4.22 $\pm$   0.02 &  20.55 $\pm$   0.06 & 0.304 & 0.875 &   4405 \\
& \multicolumn{4}{c}{ {\bf Sequence D} }   & \\
RGB &  -2.68 $\pm$   0.03 &  19.33 $\pm$   0.07 & 0.211 & 0.608 &   5588 \\
O-rich &  -3.58 $\pm$   0.04 &  21.39 $\pm$   0.10 & 0.217 & 0.752 &   3354 \\
C-rich &  -3.98 $\pm$   0.08 &  22.17 $\pm$   0.25 & 0.299 & 0.701 &    969 \\
Extreme AGB &  -4.71 $\pm$   1.40 &  23.72 $\pm$   4.28 & 0.545 & 0.414 &     18 \\
All Stars &  -4.16 $\pm$   0.02 &  23.12 $\pm$   0.04 & 0.288 & 0.863 &   9929 \\
& \multicolumn{4}{c}{ {\bf Sequence E} }  & \\
RGB &  -2.62 $\pm$   0.02 &  18.52 $\pm$   0.05 & 0.274 & 0.894 &   1718 \\
All Stars &  -2.62 $\pm$   0.02 &  18.52 $\pm$   0.05 & 0.274 & 0.894 &   1718 \\
\enddata
\tablecomments{Parameters for the linear fit to the IRAC [3.6] $\log{P}$-magnitude relation.  Identical tables for the \jmag, \hmag,  [4.5], [5.8], [8.0] and [24] bands are available online.  ``Residual" is the standard error of the residuals to the fit.  The r$^{2}$ column lists the correlation coefficient, and ``number" lists the number of sources used to derive each fit.}
\label{tab:fit}
\end{deluxetable}

\begin{deluxetable}{lccccc}
\tabletypesize{\scriptsize}
\tablecaption{\ks\ vs. $\log{\rm P}$ linear fit parameters }
\tablewidth{0pt}
\tablehead{
\colhead{Classification} & \colhead{Slope} & \colhead{Intercept} & \colhead{Residual} & \colhead{r$^{2}$} & \colhead{Number} }
\startdata
&\multicolumn{4}{c}{ {\bf Sequence 4} }   & \\
RGB &  -3.43 $\pm$   0.05 &  17.22 $\pm$   0.07 & 0.111 & 0.677 &   2054 \\
O-rich &  -3.73 $\pm$   0.04 &  17.43 $\pm$   0.06 & 0.126 & 0.867 &   1434 \\
C-rich &  -3.73 $\pm$   0.32 &  17.58 $\pm$   0.52 & 0.134 & 0.833 &     29 \\
All Stars &  -4.46 $\pm$   0.02 &  18.65 $\pm$   0.03 & 0.135 & 0.929 &   3517 \\
 &\multicolumn{4}{c}{ {\bf Sequence 3}}   & \\
RGB &  -3.12 $\pm$   0.05 &  17.36 $\pm$   0.08 & 0.117 & 0.684 &   1981 \\
O-rich &  -3.85 $\pm$   0.04 &  18.33 $\pm$   0.07 & 0.182 & 0.846 &   1707 \\
C-rich &  -3.61 $\pm$   0.19 &  17.99 $\pm$   0.35 & 0.182 & 0.875 &     56 \\
All Stars &  -4.33 $\pm$   0.02 &  19.27 $\pm$   0.03 & 0.166 & 0.932 &   3744 \\
& \multicolumn{4}{c}{ {\bf Sequence 2} }  & \\
O-rich &  -4.06 $\pm$   0.02 &  19.33 $\pm$   0.04 & 0.143 & 0.948 &   2292 \\
C-rich &  -3.69 $\pm$   0.04 &  18.61 $\pm$   0.09 & 0.155 & 0.899 &    891 \\
All Stars &  -3.93 $\pm$   0.01 &  19.10 $\pm$   0.03 & 0.149 & 0.963 &   3183 \\
& \multicolumn{4}{c}{ {\bf Sequence 1} }  & \\
O-rich &  -3.31 $\pm$   0.04 &  18.87 $\pm$   0.09 & 0.272 & 0.760 &   2218 \\
C-rich &  -3.16 $\pm$   0.04 &  18.40 $\pm$   0.11 & 0.230 & 0.744 &   1817 \\
Extreme AGB &  -2.56 $\pm$   0.35 &  17.43 $\pm$   0.91 & 0.444 & 0.147 &    312 \\
All Stars &  -3.34 $\pm$   0.02 &  18.90 $\pm$   0.05 & 0.293 & 0.848 &   4347 \\
& \multicolumn{4}{c}{ {\bf Sequence D} }   & \\
RGB &  -2.60 $\pm$   0.03 &  19.26 $\pm$   0.07 & 0.197 & 0.622 &   5611 \\
O-rich &  -3.40 $\pm$   0.03 &  21.04 $\pm$   0.09 & 0.197 & 0.767 &   3364 \\
C-rich &  -3.58 $\pm$   0.06 &  21.50 $\pm$   0.18 & 0.220 & 0.774 &    971 \\
Extreme AGB &  -4.41 $\pm$   0.64 &  24.04 $\pm$   1.96 & 0.246 & 0.748 &     18 \\
All Stars &  -3.81 $\pm$   0.01 &  22.34 $\pm$   0.04 & 0.236 & 0.873 &   9964 \\
& \multicolumn{4}{c}{ {\bf Sequence E} }  & \\
RGB &  -2.54 $\pm$   0.02 &  18.44 $\pm$   0.05 & 0.246 & 0.893 &   1706 \\
All Stars &  -2.54 $\pm$   0.02 &  18.44 $\pm$   0.05 & 0.246 & 0.893 &   1706 \\
\enddata
\tablecomments{Identical to table~\ref{tab:fit}, with magnitude measured in the \ks\ band}
\label{tab:kfit}
\end{deluxetable}

\begin{figure}
\begin{center}
\rotatebox{0}{\includegraphics*[scale=0.7]{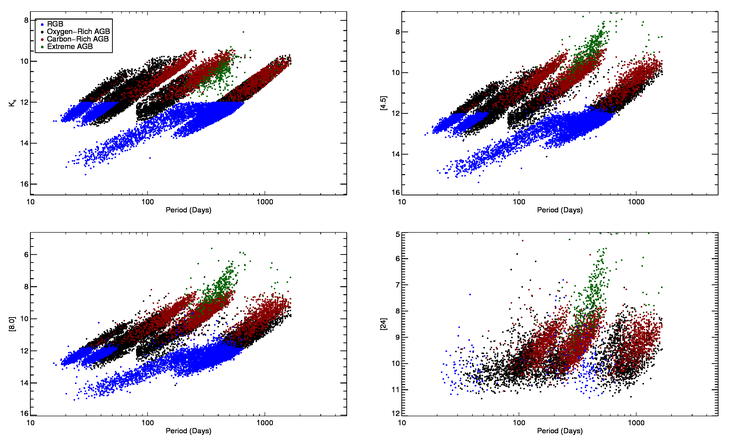}}
\caption{Four versions of the period luminosity diagram of figure~\ref{fig:pl_chem}, using different bands as the luminosity proxy. Top row, L to R: the 2MASS \ks\ band and the Spitzer IRAC $4.5$~\mic\ band from the SAGE survey.  Bottom row, L to R: the IRAC $8.0$~\mic\ band and the MIPS $24$~\mic\ band.  The O- and C-rich AGB candidates maintain the same relative position in all the bands.  The extreme sources do not look particularly bright in the \ks\ band, but further into the IR, are revealed as the brightest sources in our sample.  The RGB stars are almost entirely absent from the MIPS $24$~\mic\ diagram.}
\label{fig:4band}
\end{center}
\end{figure}

\begin{figure}
\begin{center}
\rotatebox{0}{\includegraphics*[scale=0.5]{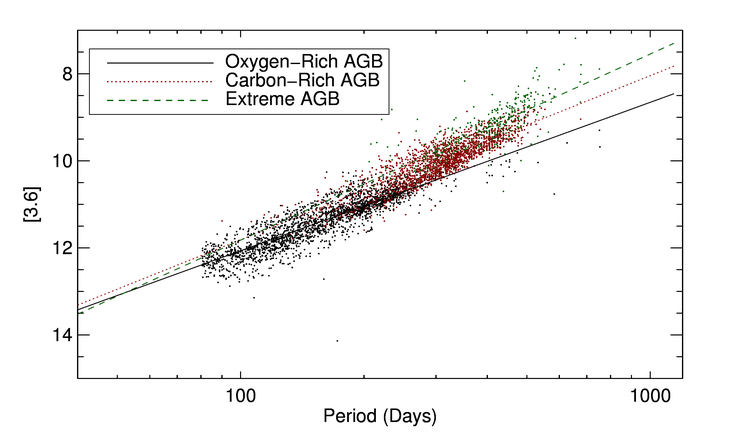}}
\caption{Linear least-squares fits to the period-[3.6] magnitude relation for the sources assigned to sequence 1, the fundamental mode pulsators.  Fits were done to each sub-group individually.  Sequence 1 is the only sequence with a significant number of extreme AGB sources on it.  Notice the smooth progression from shallow to steep slopes for the 3 different classes of AGB stars.}
\label{fig:seq1}
\end{center}
\end{figure}

\begin{figure}
\begin{center}
\rotatebox{0}{\includegraphics*[scale=0.5]{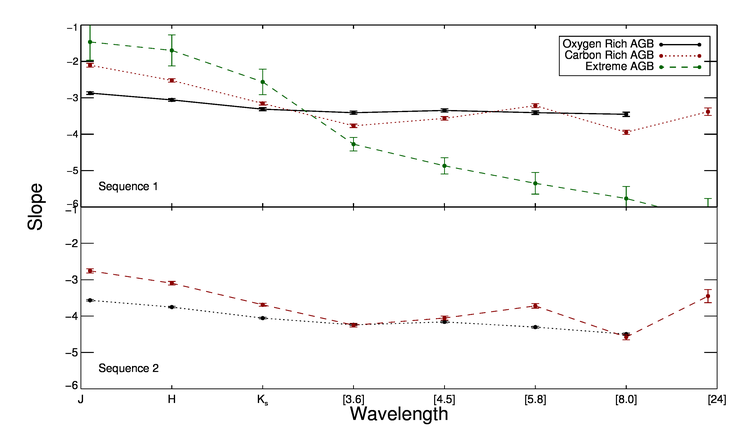}}
\caption{The slopes of the period-magnitude relations for each type of star in sequences 1 (top) and 2 (bottom), plotted for all the SAGE IR wavelengths.  Steeper slopes are more negative, and are towards the \emph{bottom} of this plot.  A steeper slope is an indication that longer period variables have brightened relative to the shorter period stars on the same sequence.}
\label{fig:1_comp}
\end{center}
\end{figure}

\begin{figure}
\begin{center}
\rotatebox{0}{\includegraphics*[scale=0.5]{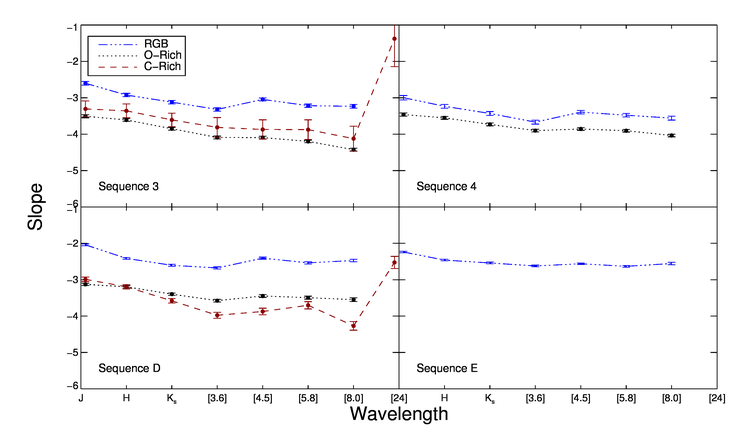}}
\caption{Similar to figure~\ref{fig:1_comp}, but showing the 4 P-L sequences not shown there.  \emph{Top row:} Left: Sequence 3  Right: Sequence 4.  \emph{Bottom row:} Left: Sequence D  Right: Sequence E}
\label{fig:allseq_comp}
\end{center}
\end{figure}

\begin{figure}
\begin{tabular}{cc}
\rotatebox{0}{\includegraphics*[scale=0.25]{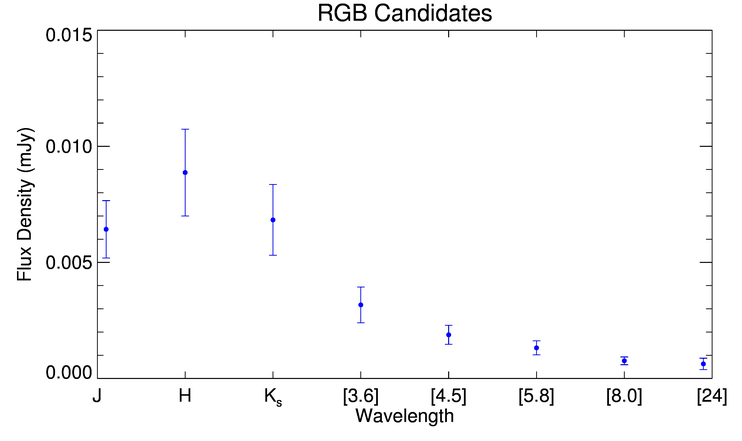}} & \rotatebox{0}{\includegraphics[scale=0.25]{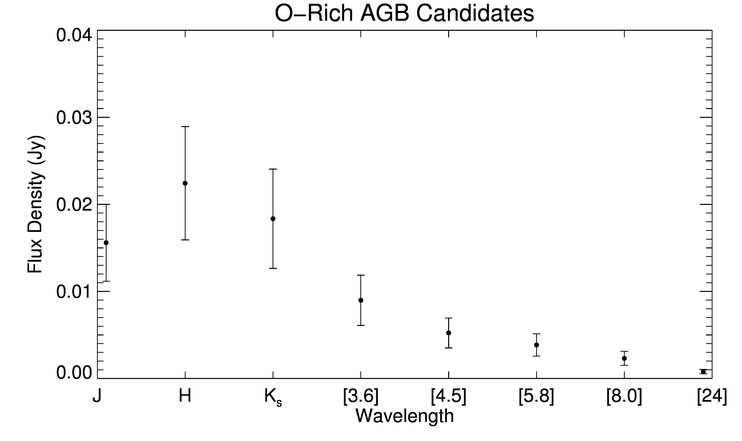}} \\
\rotatebox{0}{\includegraphics*[scale=0.25]{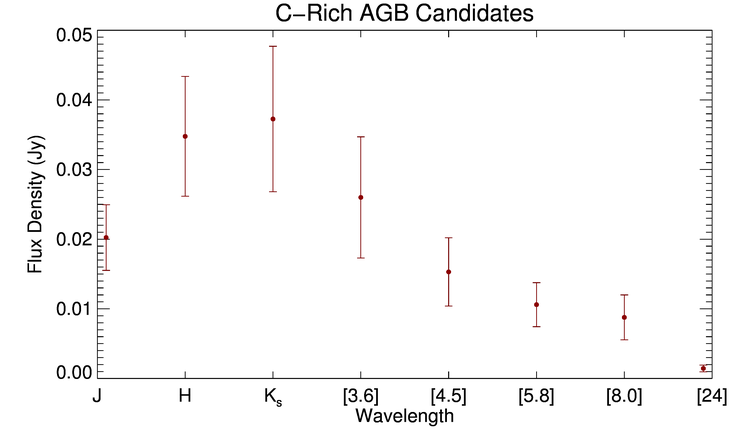}} & \rotatebox{0}{\includegraphics[scale=0.25]{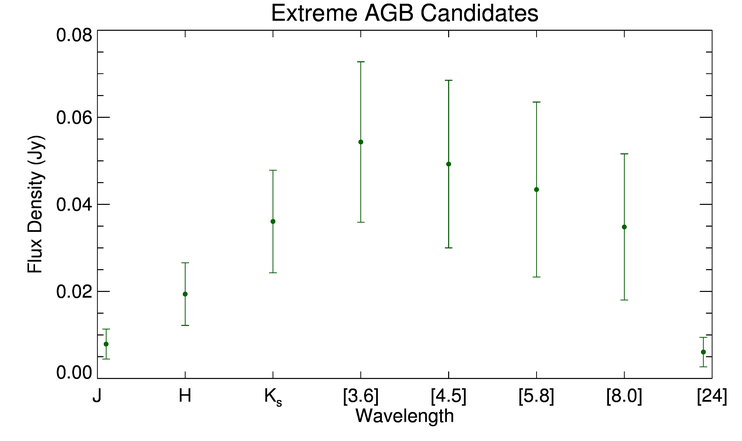}} 
\end{tabular}
\caption{Median SEDs for each of the four population classes used in this paper, first ascent RGB (top left), O-rich AGB candidates (top right), C-rich AGB candidates (bottom left) and extreme AGB candidates (bottom right).  Error bars represent the spread in the population, not measurement uncertainty.  The utilization of longer wavelength IRAC bands allows us to more accurately measure the intrinsic brightness of these sources, particularly the extreme AGB candidates, whose SEDs peak in the 3--8~\mic\ range.}
\label{fig:seds}
\end{figure}

\clearpage

\begin{figure}
\begin{center}
\rotatebox{0}{\includegraphics*[scale=0.5]{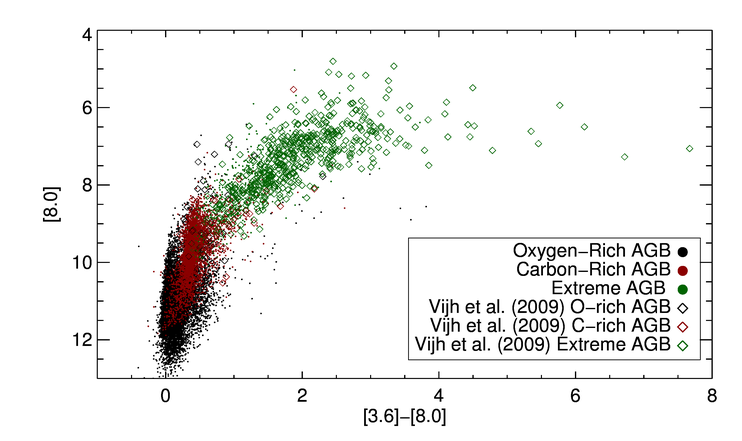}}
\caption{[8.0] vs.\ $[3.6]-[8.0]$ CMD of the AGB candidates detected by \citet{Vijh2009} overlaid atop the CMD for the sources used in this paper, detected by both SAGE and MACHO.  The sources from \citet{Vijh2009} appear as large, open diamonds.  Note the long tail of very red extreme AGB candidates $([3.6]-[8.0]) \gtrsim 4$ detected by SAGE but not MACHO.}
\label{fig:u_cmd}
\end{center}
\end{figure}

\begin{figure}
\begin{center}
\rotatebox{0}{\includegraphics*[scale=0.5]{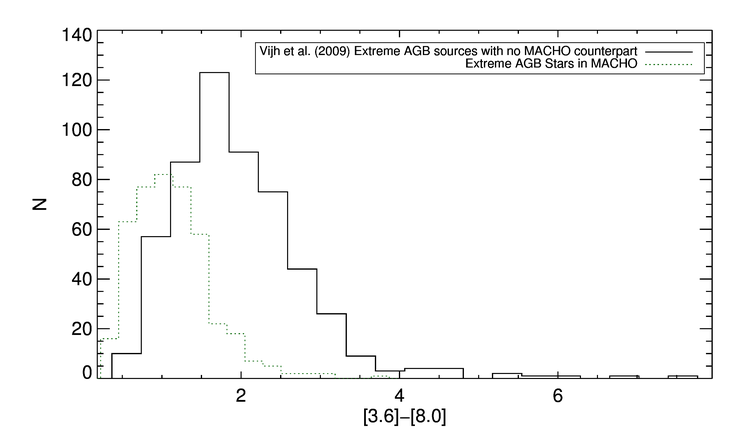}}
\caption{Histogram of $[3.6]-[8.0]$ color of the sources classified as extreme AGB candidates in this work and in \citet{Vijh2009}.  Because we require a valid MACHO (optical) detection, we miss the reddest and most enshrouded of the \citet{Vijh2009} extreme sources.}
\label{fig:u_hist}
\end{center}
\end{figure}

\begin{figure}
\begin{center}
\rotatebox{0}{\includegraphics*[scale=0.5]{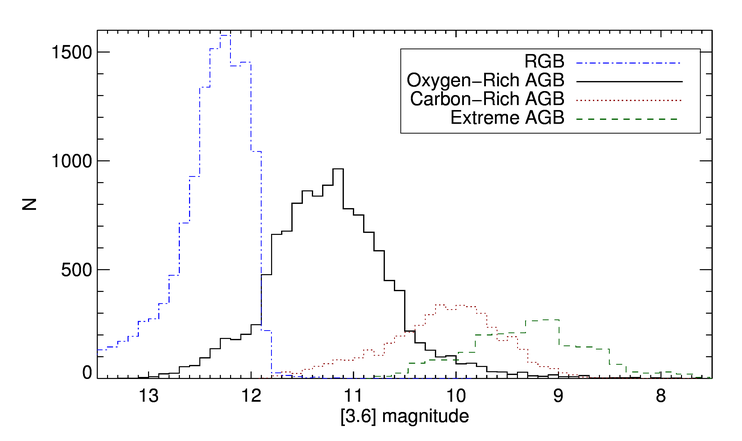}}
\caption{Histogram of IRAC $3.6$~\mic\ magnitude for the four types of sources in our sample.  The extreme AGB sample has been artificially replicated 5 times to make it visible on the same scale as the others.}
\label{fig:mag_hist}
\end{center}
\end{figure}

\begin{figure}
\begin{center}
\rotatebox{0}{\includegraphics*[scale=0.5]{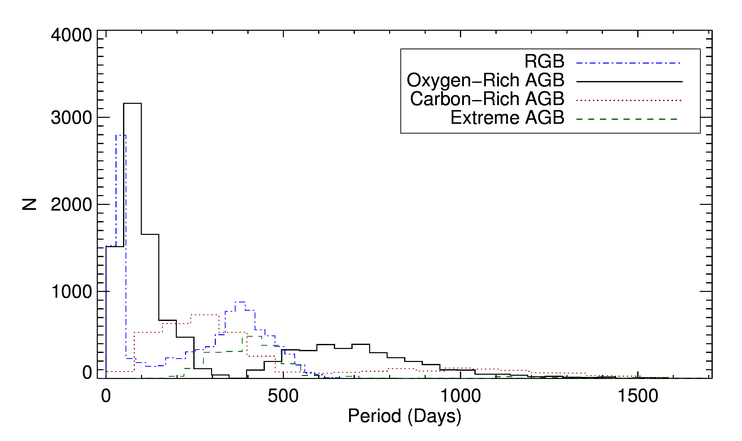}}
\caption{Histogram of MACHO primary period for the four types of sources in our sample.  The extreme AGB sample has been artificially replicated 5 times to make it visible on the same scale as the others.}
\label{fig:per_hist}
\end{center}
\end{figure}

\clearpage

\appendix  
\section{Period-Luminosity Sequence Definitions} \label{app:p_l}

\begin{figure}
\begin{center}
\rotatebox{0}{\includegraphics*[scale=0.5]{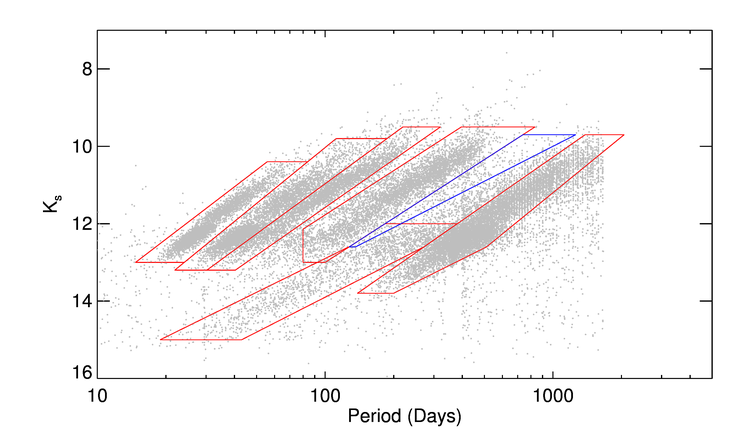}}
\caption{\ks\ band P-L diagram of our sample used in this study, showing the boundaries which define the 6 P-L sequences identified in this work.  The populations of these sources are detailed in Table~\ref{tab:pops_me}.  The boundaries are described quantitatively in table~\ref{tab:pl_def}.  The blue region in between sequences 1 and D categorizes stars with \jmag\ -- \ks\ $> 1.4$ as being on sequence 1.}
\label{fig:app_1}
\end{center}
\end{figure}

The stars in our sample are assigned to one of 6 P-L sequences using cuts in \ks-Period space, with a small subset also utilizing \jmag--\ks\ color information.  Only the primary period of these multi-periodic objects is used to establish sequence membership.  The definitions for the sequences are shown graphically in figure~\ref{fig:app_1}, and described mathematically in Table~\ref{tab:pl_def}.  Our definitions are based on those used by \citet{Fraser2008}, but are more restrictive (especially in sequence D, which exhibits a very large scatter about the P-L relation) to better trace the population density ``cores" visible to the eye.  This emphasis sacrifices some sample size, but we note that the scatter in our P-L relations (of order $\sim$0.15~mag) is dominated by the intrinsic variability of the sources and would not be significantly reduced by introducing more sources by using the secondary periods to assign the un-sequenced sources to one of the existing P-L sequences.

We note that ultimately, our P-L sequence definitions are a judgment call, and slightly different definitons used by other authors are equally valid and can affect the measured P-L fits.  In particular, the TRGB cut at a constant $\mbox{\ks} = 12.0$ leads to a sharp parallelogram shape at the brightest end of the RGB population.  These corners cause the RGB and O-rich AGB P-L fits to be less steep than the fit to all the stars in a sequence, regardless of class, which is not affected by this selection bias (Table~\ref{tab:fit}).

Beginning with these definitions of the P-L sequences, careful study of the panels of Figure~\ref{fig:4band} led us to introduce some additional modifications to properly classify a few anomalous populations.

A group of 100 C-rich AGB candidates, classified as being part of sequence 1 under criterion 1$^{b}$ in table~\ref{tab:pl_def}, was found to remain sub-luminous to the rest of sequence 1 at all wavelengths longer than \ks\ as well.   They are identified and eliminated by removing C-rich AGB candidates from sequence 1 with periods $> 300$~days and \ks\ $>-3.9*\log{\rm P}+20.17$.

Finally, we identified a population of 60 extreme AGB candidates so dust-enshrouded that at \ks\ they lay along the main body of sequence D, but at $3.6$~\mic\ and longer they were clearly part of sequence 1.  They were properly classified by adding all extreme AGB candidates assigned to sequence D or to no sequence with periods between $250$ and $700$~days and IRAC [3.6] $<10.5$ to sequence 1.

Note that both of these populations also require their members to be photometrically classified as either C-rich or extreme AGB candidates, following the definitions detailed in \S~\ref{sec:class}.



\begin{landscape}
\begin{deluxetable}{lllll}
\tabletypesize{\scriptsize}
\tablecaption{P-L Sequence Definitions}
\tablewidth{0pt}
\tablehead{Sequence & \colhead{Top} & \colhead{Bottom} & \colhead{Left} &  \colhead{Right} }
\startdata
     Sequence 1\tablenotemark{a} & \ks\ $\ge 9.5$   & \ks\ $< 13$     &  \ks\ $>-3.8*\log{\rm P}+19.37$   & \ks\ $\le -3.8*\log{\rm P}+20.6$ \\
	 Sequence 1\tablenotemark{b} &  \ks\ $\ge 9.7$ & \ks\ $< 12.6$  & \ks\ $>-3.8*\log{\rm P}+20.6$  & \ks\ $\le -3*\log{\rm P}+19$  \\
	 Sequence 2 & \ks\ $\ge 9.5$   & \ks\ $< 13.2$   & \ks\ $>-4.3*\log{\rm P}+19.57$  & \ks\ $\le -4.11*\log{\rm P}+19.8$       \\
     Sequence 3 & \ks\ $\ge 9.8$   & \ks\ $< 13.2$   & \ks\ $> -4.8*\log{\rm P}+19.63$ & \ks\ $\le -4.3*\log{\rm P}+19.57$       \\
     Sequence 4 & \ks\ $\ge 10.4$  & \ks\ $< 13$     & \ks\ $> -4.5*\log{\rm P}+18.25$ & \ks\ $\le -4.8*\log{\rm P}+19.63$      \\
     Sequence D (top) & \ks\ $ \ge 9.7$   & \ks\ $< 12.6$   & \ks\ $>-4.1*\log{\rm P}+22.58$ & \ks\ $\le -4.8*\log{\rm P}+25.6$  \\
	 Sequence D (bottom) & \ks\ $\ge 12.6$ & \ks\ $< 13.8$  & \ks\ $> -4.1*\log{\rm P}+22.58$ & \ks\ $\le -3*\log{\rm P}+20.7$    \\
     Sequence E (top)& \ks\ $> 12.0$ & \ks\ $< 12.6$ & \ks\ $\ge -3.8*\log{\rm P}+20.6$ & \ks\ $< -4.1*\log{\rm P}+22.58$ \\
     Sequence E (bottom)& \ks\ $\ge 12.6$  & \ks\ $< 15$   & \ks\ $>-2.9*\log{\rm P}+18.7$ & \ks\ $\le -3*\log{\rm P}+19.9$
\enddata
\tablenotetext{a}{There is an additional constraint that stars in sequence 1 must have a period greater than $80$~days.  This requirement only affects a small number of stars and serves to prevent these outliers from skewing the P-L fits.}
\tablenotetext{b}{This region, shown in blue in Figure~\ref{fig:app_1} uses an additional requirement that \jmag--\ks $>1.4$ to identify enshrouded stars (many of which are extreme AGB candidates) that would otherwise be omitted.}
\tablecomments{Mathematical description of the boundaries shown in Figure~\ref{fig:app_1}.}
\label{tab:pl_def}
\end{deluxetable}
\end{landscape}

\section{Mid-IR Period-Magnitude Relations} \label{app:fits}
\begin{deluxetable}{lccccc}
\tabletypesize{\scriptsize}
\tablecaption{J vs. $\log{\rm P}$ linear fit parameters }
\tablewidth{0pt}
\tablehead{
\colhead{Classification} & \colhead{Slope} & \colhead{Intercept} & \colhead{Residual}  & \colhead{r$^{2}$}& \colhead{Number} }
\startdata
&\multicolumn{4}{c}{ {\bf Sequence 4} }   & \\
RGB &  -3.00 $\pm$   0.06 &  17.65 $\pm$   0.08 & 0.126 & 0.558 &   2050 \\
O-rich &  -3.46 $\pm$   0.04 &  18.08 $\pm$   0.07 & 0.142 & 0.814 &   1434 \\
C-rich &  -3.74 $\pm$   0.33 &  18.95 $\pm$   0.53 & 0.128 & 0.844 &     26 \\
All Stars &  -4.16 $\pm$   0.02 &  19.27 $\pm$   0.04 & 0.153 & 0.899 &   3510 \\
 &\multicolumn{4}{c}{ {\bf Sequence 3}}   & \\
RGB &  -2.60 $\pm$   0.05 &  17.56 $\pm$   0.08 & 0.125 & 0.579 &   1979 \\
O-rich &  -3.51 $\pm$   0.04 &  18.82 $\pm$   0.08 & 0.195 & 0.797 &   1707 \\
C-rich &  -3.31 $\pm$   0.22 &  18.79 $\pm$   0.42 & 0.214 & 0.823 &     51 \\
All Stars &  -3.91 $\pm$   0.02 &  19.62 $\pm$   0.03 & 0.181 & 0.906 &   3737 \\
& \multicolumn{4}{c}{ {\bf Sequence 2} }  & \\
O-rich &  -3.57 $\pm$   0.02 &  19.51 $\pm$   0.04 & 0.154 & 0.927 &   2290 \\
C-rich &  -2.76 $\pm$   0.06 &  18.15 $\pm$   0.12 & 0.220 & 0.719 &    887 \\
All Stars &  -2.96 $\pm$   0.02 &  18.43 $\pm$   0.04 & 0.223 & 0.887 &   3177 \\
& \multicolumn{4}{c}{ {\bf Sequence 1} }  & \\
O-rich &  -2.87 $\pm$   0.04 &  18.99 $\pm$   0.08 & 0.252 & 0.747 &   2219 \\
C-rich &  -2.10 $\pm$   0.05 &  17.46 $\pm$   0.13 & 0.307 & 0.460 &   1812 \\
Extreme AGB &  -1.47 $\pm$   0.50 &  17.19 $\pm$   1.32 & 0.767 & 0.027 &    311 \\
All Stars &  -1.95 $\pm$   0.03 &  17.06 $\pm$   0.06 & 0.500 & 0.539 &   4342 \\
& \multicolumn{4}{c}{ {\bf Sequence D} }   & \\
RGB &  -2.04 $\pm$   0.03 &  18.81 $\pm$   0.07 & 0.196 & 0.515 &   5590 \\
O-rich &  -3.13 $\pm$   0.03 &  21.40 $\pm$   0.09 & 0.190 & 0.749 &   3364 \\
C-rich &  -2.99 $\pm$   0.06 &  21.25 $\pm$   0.19 & 0.240 & 0.700 &    965 \\
Extreme AGB &  -2.11 $\pm$   1.81 &  19.40 $\pm$   5.55 & 0.600 & 0.088 &     16 \\
All Stars &  -3.15 $\pm$   0.01 &  21.64 $\pm$   0.04 & 0.248 & 0.828 &   9935 \\
& \multicolumn{4}{c}{ {\bf Sequence E} }  & \\
RGB &  -2.24 $\pm$   0.02 &  18.73 $\pm$   0.04 & 0.236 & 0.878 &   1703 \\
All Stars &  -2.24 $\pm$   0.02 &  18.73 $\pm$   0.04 & 0.236 & 0.878 &   1703 \\
\enddata
\tablecomments{Identical to table~\ref{tab:fit}, with magnitude measured in the J band}
\end{deluxetable}

\begin{deluxetable}{lccccc}
\tabletypesize{\scriptsize}
\tablecaption{H vs. $\log{\rm P}$ linear fit parameters }
\tablewidth{0pt}
\tablehead{
\colhead{Classification} & \colhead{Slope} & \colhead{Intercept} & \colhead{Residual}  & \colhead{r$^{2}$}& \colhead{Number} }
\startdata
&\multicolumn{4}{c}{ {\bf Sequence 4} }   & \\
RGB &  -3.23 $\pm$   0.05 &  17.14 $\pm$   0.07 & 0.112 & 0.648 &   2048 \\
O-rich &  -3.55 $\pm$   0.04 &  17.37 $\pm$   0.07 & 0.132 & 0.842 &   1432 \\
C-rich &  -3.37 $\pm$   0.33 &  17.32 $\pm$   0.53 & 0.145 & 0.798 &     28 \\
All Stars &  -4.26 $\pm$   0.02 &  18.56 $\pm$   0.03 & 0.139 & 0.919 &   3508 \\
 &\multicolumn{4}{c}{ {\bf Sequence 3}}   & \\
RGB &  -2.92 $\pm$   0.05 &  17.24 $\pm$   0.07 & 0.115 & 0.668 &   1975 \\
O-rich &  -3.61 $\pm$   0.04 &  18.14 $\pm$   0.08 & 0.188 & 0.818 &   1704 \\
C-rich &  -3.36 $\pm$   0.19 &  17.90 $\pm$   0.36 & 0.190 & 0.863 &     52 \\
All Stars &  -4.06 $\pm$   0.02 &  19.02 $\pm$   0.03 & 0.169 & 0.923 &   3731 \\
& \multicolumn{4}{c}{ {\bf Sequence 2} }  & \\
O-rich &  -3.75 $\pm$   0.02 &  19.00 $\pm$   0.04 & 0.145 & 0.941 &   2291 \\
C-rich &  -3.10 $\pm$   0.05 &  17.83 $\pm$   0.11 & 0.187 & 0.814 &    888 \\
All Stars &  -3.35 $\pm$   0.02 &  18.28 $\pm$   0.03 & 0.181 & 0.935 &   3179 \\
& \multicolumn{4}{c}{ {\bf Sequence 1} }  & \\
O-rich &  -3.06 $\pm$   0.04 &  18.56 $\pm$   0.08 & 0.267 & 0.753 &   2218 \\
C-rich &  -2.52 $\pm$   0.05 &  17.40 $\pm$   0.12 & 0.268 & 0.605 &   1814 \\
Extreme AGB &  -1.70 $\pm$   0.43 &  16.30 $\pm$   1.12 & 0.620 & 0.048 &    314 \\
All Stars &  -2.53 $\pm$   0.02 &  17.45 $\pm$   0.06 & 0.398 & 0.720 &   4346 \\
& \multicolumn{4}{c}{ {\bf Sequence D} }   & \\
RGB &  -2.41 $\pm$   0.03 &  18.95 $\pm$   0.07 & 0.194 & 0.596 &   5590 \\
O-rich &  -3.19 $\pm$   0.03 &  20.72 $\pm$   0.09 & 0.189 & 0.757 &   3361 \\
C-rich &  -3.19 $\pm$   0.06 &  20.83 $\pm$   0.18 & 0.220 & 0.745 &    967 \\
Extreme AGB &  -2.30 $\pm$   1.07 &  18.57 $\pm$   3.27 & 0.375 & 0.249 &     16 \\
All Stars &  -3.46 $\pm$   0.01 &  21.60 $\pm$   0.04 & 0.231 & 0.860 &   9934 \\
& \multicolumn{4}{c}{ {\bf Sequence E} }  & \\
RGB &  -2.46 $\pm$   0.02 &  18.43 $\pm$   0.04 & 0.237 & 0.896 &   1700 \\
All Stars &  -2.46 $\pm$   0.02 &  18.43 $\pm$   0.04 & 0.237 & 0.896 &   1700 \\
\enddata
\tablecomments{Identical to table~\ref{tab:fit}, with magnitude measured in the H band}
\end{deluxetable}

\begin{deluxetable}{lccccc}
\tabletypesize{\scriptsize}
\tablecaption{[4.5] vs. $\log{\rm P}$ linear fit parameters }
\tablewidth{0pt}
\tablehead{
\colhead{Classification} & \colhead{Slope} & \colhead{Intercept} & \colhead{Residual}  & \colhead{r$^{2}$}& \colhead{Number} }
\startdata
&\multicolumn{4}{c}{ {\bf Sequence 4} }   & \\
RGB &  -3.39 $\pm$   0.05 &  17.18 $\pm$   0.07 & 0.110 & 0.711 &   2065 \\
O-rich &  -3.86 $\pm$   0.04 &  17.66 $\pm$   0.06 & 0.121 & 0.885 &   1434 \\
C-rich &  -4.14 $\pm$   0.35 &  18.17 $\pm$   0.57 & 0.142 & 0.836 &     29 \\
All Stars &  -4.42 $\pm$   0.02 &  18.61 $\pm$   0.03 & 0.128 & 0.938 &   3528 \\
 &\multicolumn{4}{c}{ {\bf Sequence 3}}   & \\
RGB &  -3.04 $\pm$   0.04 &  17.23 $\pm$   0.07 & 0.104 & 0.728 &   1992 \\
O-rich &  -4.09 $\pm$   0.04 &  18.77 $\pm$   0.07 & 0.172 & 0.877 &   1706 \\
C-rich &  -3.87 $\pm$   0.27 &  18.26 $\pm$   0.51 & 0.246 & 0.796 &     56 \\
All Stars &  -4.36 $\pm$   0.02 &  19.31 $\pm$   0.03 & 0.155 & 0.943 &   3755 \\
& \multicolumn{4}{c}{ {\bf Sequence 2} }  & \\
O-rich &  -4.16 $\pm$   0.02 &  19.43 $\pm$   0.04 & 0.154 & 0.936 &   2290 \\
C-rich &  -4.05 $\pm$   0.05 &  18.99 $\pm$   0.12 & 0.204 & 0.862 &    891 \\
All Stars &  -4.44 $\pm$   0.02 &  19.93 $\pm$   0.03 & 0.180 & 0.955 &   3181 \\
& \multicolumn{4}{c}{ {\bf Sequence 1} }  & \\
O-rich &  -3.35 $\pm$   0.05 &  18.80 $\pm$   0.10 & 0.270 & 0.708 &   2227 \\
C-rich &  -3.56 $\pm$   0.05 &  18.96 $\pm$   0.12 & 0.265 & 0.750 &   1816 \\
Extreme AGB &  -4.87 $\pm$   0.22 &  21.55 $\pm$   0.58 & 0.402 & 0.560 &    371 \\
All Stars &  -4.16 $\pm$   0.03 &  20.47 $\pm$   0.06 & 0.367 & 0.854 &   4414 \\
& \multicolumn{4}{c}{ {\bf Sequence D} }   & \\
RGB &  -2.41 $\pm$   0.03 &  18.72 $\pm$   0.07 & 0.214 & 0.562 &   5610 \\
O-rich &  -3.45 $\pm$   0.04 &  21.14 $\pm$   0.11 & 0.242 & 0.703 &   3361 \\
C-rich &  -3.87 $\pm$   0.09 &  21.93 $\pm$   0.27 & 0.338 & 0.646 &    969 \\
Extreme AGB &  -5.24 $\pm$   1.95 &  25.04 $\pm$   5.94 & 0.690 & 0.312 &     18 \\
All Stars &  -4.01 $\pm$   0.02 &  22.82 $\pm$   0.05 & 0.316 & 0.838 &   9958 \\
& \multicolumn{4}{c}{ {\bf Sequence E} }  & \\
RGB &  -2.56 $\pm$   0.02 &  18.46 $\pm$   0.05 & 0.275 & 0.891 &   1727 \\
All Stars &  -2.56 $\pm$   0.02 &  18.46 $\pm$   0.05 & 0.275 & 0.891 &   1727 \\
\enddata
\tablecomments{Identical to table~\ref{tab:fit}, with magnitude measured in the [4.5] band}
\end{deluxetable}

\begin{deluxetable}{lccccc}
\tabletypesize{\scriptsize}
\tablecaption{[5.8] vs. $\log{\rm P}$ linear fit parameters }
\tablewidth{0pt}
\tablehead{
\colhead{Classification} & \colhead{Slope} & \colhead{Intercept} & \colhead{Residual}  & \colhead{r$^{2}$}& \colhead{Number} }
\startdata
&\multicolumn{4}{c}{ {\bf Sequence 4} }   & \\
RGB &  -3.48 $\pm$   0.05 &  17.19 $\pm$   0.07 & 0.105 & 0.712 &   2063 \\
O-rich &  -3.90 $\pm$   0.04 &  17.61 $\pm$   0.06 & 0.118 & 0.891 &   1432 \\
C-rich &  -4.27 $\pm$   0.34 &  18.25 $\pm$   0.54 & 0.134 & 0.858 &     29 \\
All Stars &  -4.52 $\pm$   0.02 &  18.63 $\pm$   0.03 & 0.126 & 0.942 &   3524 \\
 &\multicolumn{4}{c}{ {\bf Sequence 3}}   & \\
RGB &  -3.22 $\pm$   0.04 &  17.40 $\pm$   0.07 & 0.112 & 0.723 &   1985 \\
O-rich &  -4.20 $\pm$   0.04 &  18.83 $\pm$   0.07 & 0.170 & 0.884 &   1705 \\
C-rich &  -3.88 $\pm$   0.27 &  18.11 $\pm$   0.51 & 0.248 & 0.794 &     56 \\
All Stars &  -4.53 $\pm$   0.02 &  19.47 $\pm$   0.03 & 0.158 & 0.946 &   3747 \\
& \multicolumn{4}{c}{ {\bf Sequence 2} }  & \\
O-rich &  -4.30 $\pm$   0.02 &  19.56 $\pm$   0.04 & 0.160 & 0.935 &   2288 \\
C-rich &  -3.72 $\pm$   0.06 &  18.16 $\pm$   0.14 & 0.234 & 0.794 &    887 \\
All Stars &  -4.45 $\pm$   0.02 &  19.80 $\pm$   0.04 & 0.190 & 0.948 &   3175 \\
& \multicolumn{4}{c}{ {\bf Sequence 1} }  & \\
O-rich &  -3.41 $\pm$   0.05 &  18.77 $\pm$   0.10 & 0.287 & 0.697 &   2223 \\
C-rich &  -3.22 $\pm$   0.05 &  18.04 $\pm$   0.13 & 0.308 & 0.664 &   1812 \\
Extreme AGB &  -5.35 $\pm$   0.30 &  22.44 $\pm$   0.77 & 0.509 & 0.470 &    369 \\
All Stars &  -4.02 $\pm$   0.03 &  20.04 $\pm$   0.07 & 0.439 & 0.820 &   4404 \\
& \multicolumn{4}{c}{ {\bf Sequence D} }   & \\
RGB &  -2.54 $\pm$   0.03 &  18.94 $\pm$   0.08 & 0.230 & 0.557 &   5594 \\
O-rich &  -3.49 $\pm$   0.04 &  21.13 $\pm$   0.12 & 0.259 & 0.690 &   3355 \\
C-rich &  -3.70 $\pm$   0.10 &  21.27 $\pm$   0.29 & 0.374 & 0.589 &    970 \\
Extreme AGB &  -5.43 $\pm$   2.30 &  25.23 $\pm$   7.02 & 0.781 & 0.258 &     18 \\
All Stars &  -4.20 $\pm$   0.02 &  23.18 $\pm$   0.05 & 0.333 & 0.841 &   9937 \\
& \multicolumn{4}{c}{ {\bf Sequence E} }  & \\
RGB &  -2.63 $\pm$   0.02 &  18.52 $\pm$   0.05 & 0.292 & 0.877 &   1711 \\
All Stars &  -2.63 $\pm$   0.02 &  18.52 $\pm$   0.05 & 0.292 & 0.877 &   1711 \\
\enddata
\tablecomments{Identical to table~\ref{tab:fit}, with magnitude measured in the [5.8] band}
\end{deluxetable}

\begin{deluxetable}{lccccc}
\tabletypesize{\scriptsize}
\tablecaption{[8.0] vs. $\log{\rm P}$ linear fit parameters }
\tablewidth{0pt}
\tablehead{
\colhead{Classification} & \colhead{Slope} & \colhead{Intercept} & \colhead{Residual}  & \colhead{r$^{2}$}& \colhead{Number} }
\startdata
&\multicolumn{4}{c}{ {\bf Sequence 4} }   & \\
RGB &  -3.56 $\pm$   0.05 &  17.29 $\pm$   0.08 & 0.122 & 0.682 &   2044 \\
O-rich &  -4.04 $\pm$   0.04 &  17.79 $\pm$   0.06 & 0.123 & 0.892 &   1429 \\
C-rich &  -4.50 $\pm$   0.33 &  18.58 $\pm$   0.54 & 0.140 & 0.871 &     29 \\
All Stars &  -4.62 $\pm$   0.02 &  18.76 $\pm$   0.03 & 0.136 & 0.940 &   3502 \\
 &\multicolumn{4}{c}{ {\bf Sequence 3}}   & \\
RGB &  -3.23 $\pm$   0.05 &  17.40 $\pm$   0.08 & 0.131 & 0.684 &   1978 \\
O-rich &  -4.43 $\pm$   0.04 &  19.19 $\pm$   0.07 & 0.179 & 0.885 &   1704 \\
C-rich &  -4.12 $\pm$   0.34 &  18.39 $\pm$   0.65 & 0.324 & 0.734 &     55 \\
All Stars &  -4.68 $\pm$   0.02 &  19.69 $\pm$   0.03 & 0.174 & 0.942 &   3738 \\
& \multicolumn{4}{c}{ {\bf Sequence 2} }  & \\
O-rich &  -4.49 $\pm$   0.03 &  19.83 $\pm$   0.05 & 0.198 & 0.911 &   2284 \\
C-rich &  -4.58 $\pm$   0.08 &  19.64 $\pm$   0.17 & 0.275 & 0.800 &    889 \\
All Stars &  -5.05 $\pm$   0.02 &  20.81 $\pm$   0.05 & 0.250 & 0.933 &   3173 \\
& \multicolumn{4}{c}{ {\bf Sequence 1} }  & \\
O-rich &  -3.45 $\pm$   0.06 &  18.74 $\pm$   0.13 & 0.344 & 0.612 &   2221 \\
C-rich &  -3.95 $\pm$   0.06 &  19.40 $\pm$   0.14 & 0.317 & 0.716 &   1816 \\
Extreme AGB &  -5.77 $\pm$   0.33 &  23.09 $\pm$   0.86 & 0.546 & 0.451 &    371 \\
All Stars &  -4.68 $\pm$   0.03 &  21.28 $\pm$   0.08 & 0.473 & 0.812 &   4408 \\
& \multicolumn{4}{c}{ {\bf Sequence D} }   & \\
RGB &  -2.47 $\pm$   0.03 &  18.74 $\pm$   0.09 & 0.265 & 0.479 &   5561 \\
O-rich &  -3.55 $\pm$   0.05 &  21.19 $\pm$   0.14 & 0.331 & 0.605 &   3361 \\
C-rich &  -4.27 $\pm$   0.12 &  22.61 $\pm$   0.35 & 0.433 & 0.575 &    969 \\
Extreme AGB &  -5.74 $\pm$   2.44 &  25.72 $\pm$   7.43 & 0.836 & 0.257 &     18 \\
All Stars &  -4.50 $\pm$   0.02 &  23.93 $\pm$   0.06 & 0.411 & 0.808 &   9909 \\
& \multicolumn{4}{c}{ {\bf Sequence E} }  & \\
RGB &  -2.55 $\pm$   0.03 &  18.29 $\pm$   0.07 & 0.349 & 0.787 &   1631 \\
All Stars &  -2.55 $\pm$   0.03 &  18.29 $\pm$   0.07 & 0.349 & 0.787 &   1631 \\
\enddata
\tablecomments{Identical to table~\ref{tab:fit}, with magnitude measured in the [8.0] band}
\end{deluxetable}

\begin{deluxetable}{lccccc}
\tabletypesize{\scriptsize}
\tablecaption{[24] vs. $\log{\rm P}$ linear fit parameters }
\tablewidth{0pt}
\tablehead{
\colhead{Classification} & \colhead{Slope} & \colhead{Intercept} & \colhead{Residual}  & \colhead{r$^{2}$}& \colhead{Number} }
\startdata
&\multicolumn{4}{c}{ {\bf Sequence 4} }   & \\
RGB &  -3.21 $\pm$   2.25 &  14.72 $\pm$   3.24 & 0.591 & 0.061 &     33 \\
O-rich &  -0.79 $\pm$   0.27 &  11.57 $\pm$   0.45 & 0.344 & 0.048 &    177 \\
All Stars &  -0.25 $\pm$   0.23 &  10.58 $\pm$   0.39 & 0.416 & 0.005 &    213 \\
 &\multicolumn{4}{c}{ {\bf Sequence 3}}   & \\
RGB &   3.71 $\pm$   1.93 &   4.15 $\pm$   3.11 & 0.666 & 0.091 &     39 \\
O-rich &  -3.02 $\pm$   0.17 &  15.82 $\pm$   0.33 & 0.429 & 0.401 &    471 \\
C-rich &  -1.38 $\pm$   0.77 &  12.20 $\pm$   1.52 & 0.449 & 0.113 &     27 \\
All Stars &  -1.97 $\pm$   0.14 &  13.69 $\pm$   0.28 & 0.497 & 0.257 &    538 \\
& \multicolumn{4}{c}{ {\bf Sequence 2} }  & \\
O-rich &  -1.45 $\pm$   0.18 &  12.75 $\pm$   0.37 & 0.537 & 0.079 &    731 \\
C-rich &  -3.45 $\pm$   0.18 &  16.78 $\pm$   0.39 & 0.448 & 0.328 &    758 \\
All Stars &  -2.84 $\pm$   0.11 &  15.49 $\pm$   0.22 & 0.505 & 0.328 &   1489 \\
& \multicolumn{4}{c}{ {\bf Sequence 1} }  & \\
O-rich &  -0.61 $\pm$   0.23 &  11.16 $\pm$   0.54 & 0.701 & 0.008 &    800 \\
C-rich &  -3.38 $\pm$   0.10 &  17.67 $\pm$   0.25 & 0.416 & 0.410 &   1587 \\
Extreme AGB &  -6.34 $\pm$   0.58 &  23.98 $\pm$   1.50 & 0.752 & 0.249 &    366 \\
All Stars &  -3.01 $\pm$   0.10 &  16.65 $\pm$   0.24 & 0.728 & 0.254 &   2753 \\
& \multicolumn{4}{c}{ {\bf Sequence D} }   & \\
RGB &   0.07 $\pm$   0.71 &   9.70 $\pm$   1.85 & 0.679 & 0.000 &    129 \\
O-rich &  -0.65 $\pm$   0.15 &  11.63 $\pm$   0.43 & 0.651 & 0.016 &   1240 \\
C-rich &  -2.53 $\pm$   0.17 &  16.84 $\pm$   0.50 & 0.480 & 0.227 &    789 \\
Extreme AGB &  -3.04 $\pm$   4.76 &  16.39 $\pm$  14.65 & 0.990 & 0.028 &     16 \\
All Stars &  -1.43 $\pm$   0.09 &  13.75 $\pm$   0.26 & 0.649 & 0.107 &   2174 \\
& \multicolumn{4}{c}{ {\bf Sequence E} }  & \\
RGB &  -1.26 $\pm$   0.71 &  11.89 $\pm$   1.58 & 1.062 & 0.084 &     36 \\
All Stars &  -1.26 $\pm$   0.71 &  11.89 $\pm$   1.58 & 1.062 & 0.084 &     36 \\
\enddata
\tablecomments{Identical to table~\ref{tab:fit}, with magnitude measured in the [24] band}
\end{deluxetable}

\end{document}